\documentclass[a4paper,fleqn,usenatbib]{mnras}

\usepackage{newtxtext,newtxmath,times}

\usepackage[T1]{fontenc}
\usepackage{ae,aecompl}

\usepackage{graphicx}	
\usepackage{amsmath}	
\usepackage{amssymb}	
\usepackage{array}

\newcommand{\rsun}{R$_\odot$}
\newcommand{\msun}{{M$_\odot$}}
\newcolumntype{x}[1]{>{\centering\arraybackslash}p{#1}}

\title[Extending common envelope simulations]{Extending Common Envelope Simulations from Roche Lobe Overflow to the Nebular Phase}
\author[T. A. Reichardt et al.]{Thomas A. Reichardt$^{1,2}$\thanks{E-mail: thomas.reichardt@students.mq.edu.au},
Orsola De Marco$^{1,2}$, 
Roberto Iaconi$^{1,2}$,
\newauthor Christopher A. Tout$^{3}$ 
and Daniel J. Price$^{4}$ \\
$^{1}$Department of Physics and Astronomy, Macquarie University, Sydney, NSW 2109, Australia \\
$^{2}$Astronomy, Astrophysics and Astrophotonics Research Centre, Macquarie University, Sydney, NSW 2109, Australia \\
$^{3}$Institute of Astronomy, University of Cambridge, Madingley Road, Cambridge CB3 0HA, UK \\
$^{4}$Monash Centre for Astrophysics and School of Physics and Astronomy, Monash University, VIC 3800, Australia
}

\date{Accepted XXX. Received YYY; in original form ZZZ}

\pubyear{2018}

\begin{document}
\label{firstpage}
\pagerange{\pageref{firstpage}--\pageref{lastpage}}
\maketitle

\begin{abstract}
We have simulated a common envelope interaction of a 0.88-\msun, 90-\rsun, red giant branch star and a 0.6-\msun, compact companion with the smoothed particle hydrodynamics code, \textsc{phantom}, from the beginning of the Roche lobe overflow phase to the beginning of the self-regulated inspiral, using three different resolutions. 
The duration of the Roche lobe overflow phase is resolution dependent and would lengthen with increased resolution beyond the $\sim$20 years observed, while the inspiral phase and the post-common envelope separation are largely independent of resolution. 
Mass transfer rates through the Lagrangian points drive the orbital evolution during the Roche lobe overflow phase, as predicted analytically. The absolute mass transfer rate is resolution dependent, but always within an order of magnitude of the analytical value. Similarly, the gravitational drag in the simulations is close to the analytical approximation. This gives us confidence that simulations approximate reality. The $L_2$ and $L_3$ outflow observed during Roche lobe overflow remains bound, forming a circumbinary disk that is largely disrupted by the common envelope ejection. However, a longer phase of Roche lobe overflow and weaker common envelope ejection typical of a more stable binary may result in a surviving circumbinary disk. 
Finally, we examine the density distribution resulting from the interaction for simulations that include or omit the phase of Roche lobe overflow. We conclude that the degree of stability of the Roche lobe phase may modulate the shape of the subsequent planetary nebula, explaining the wide range of post-common envelope planetary nebula shapes observed.
\end{abstract}




\section{Introduction}

The common envelope (CE) interaction is a phase of binary stellar evolution invoked to describe a dramatic decrease in the orbital separation of two stars, coupled with the ejection of a large amount of gas. The idea is attributed to \citet[][by whom private communications with Ostriker and Webbink are also credited]{paczynski1976common}, who was attempting to explain the origin of some compact, evolved binaries. As the primary, more massive star in an intermediate separation binary system swells to become a red giant, it transfers mass to its companion in a phase of Roche lobe overflow. If the mass transfer is unstable, the companion is likely swallowed by the expanding outer atmosphere of the primary. What follows is a phase of dramatic reduction in the orbital separation on a dynamical time-scale, after which the envelope is presumably ejected to reveal a compact binary composed of the companion and the core of the giant primary. If the envelope is not ejected then it is expected that the stars will eventually merge. 

Understanding this interaction is critical to the interpretation of binary phenomena, from intermediate luminosity red transients \citep[e.g.,][]{blagorodnova2017common} to determining the nature of the progenitor of type Ia supernovae \citep[e.g.,][]{toonen2013effect}, and to illuminating the evolutionary path that leads to compact neutron star and black hole systems that can merge by emission of gravitational waves \citep[e.g.,][]{abbott2016observation}. See \citet{de2017dawes} for a review.

Several observed objects carry characteristics distinctive of the common envelope interaction and have been interpreted as common envelope mergers. While some have been known for a while (V838~Monocerotis, \citealt{bond2003energetic}), many more are being discovered now, with the intensification of time-resolved observations. V1309~Sco \citep{tylenda2011v1309} was interpreted as the merging of a solar-mass subgiant with a much less massive companion. M31-2015LRN \citep{macleod2017lessons}, and possibly V838~Mon, were likely slightly more massive, in the range of 3-5.5\,\msun. M101-OT \citep{blagorodnova2017common} possibly came from an 18\,\msun\ progenitor, while NGC~4490-OT \citep{smith2016massive} might have derived from an even more massive star. This flurry of observations promises to place much needed constraints on common envelope simulations, which in turn can start to give a physical explanations to these phenomena \citep{galaviz2017common}.

Simulations of the common envelope utilising a variety of codes have, with mild success, reproduced observable post-CE system parameters. A non-exhaustive list of recent works includes the AMR simulations \citep{ricker2012amr}, comparisons of SPH and grid techniques \citep{passy2012simulating, iaconi2017effect}, SPH simulations examining the inclusion of recombination energy \citep{nandez2015recombination} and moving-mesh simulations \citep{ohlmann2016hydrodynamic}. Most simulations model lower-mass red giant branch stars and thereby only cover a limited section of the parameter space. In addition, simulations typically start with the companion very close to the surface of the star and stop as soon as the orbital distance has stabilised. This leaves open the question of whether the phases surrounding the dynamic inspiral have an effect on the inspiral itself, on the parameters of the post-CE binaries, including the ejecta or nebula that may derive from it, and on the light properties near the time of the outburst.

The effect of starting common envelope simulations at primary Roche lobe contact was investigated by \citet{iaconi2017effect}, who concluded that including the unstable Roche lobe overflow phase leads to a marginally wider final separation. They also showed that starting the simulations at a wider separation does not increase the amount of unbound common envelope mass\footnote{Envelope unbinding remains an issue, although some solutions or partial solutions have started appearing in the literature. Recent discussion has turned towards the ability of recombination energy to act as a source of energy \citep{han1994possible}, allowing further envelope unbinding (see, for example, \citealt{ivanova2018use} and the counter arguments of \citealt{grichener2018limited} and \citealt{soker2018radiating}).}. However, they did not investigate the outflow, nor the details of the early, pre-dynamical, inspiral. In one of their simulations, carried out with the SPH code \textsc{phantom} \citep{price2017phantom}, they noticed that the relatively long phase of Roche lobe overflow, before the fast inspiral, resulted in the ejection of some mass from the second and third Lagrangian points, $L_2$ and $L_3$. This raised the suspicion that the modelled common envelope ejecta would differ from when the simulation starts at smaller orbital separations, something that could impact studies of the nebular shapes resulting from common envelope ejections \citep{frank2018planetary,garcia2018common}. 
 
Observations already indicate that phases just before the dynamical common envelope may play a role in the overall outcome of the interaction. For example, V1309~Scorpii was captured in the OGLE-III and OGLE-IV fields between 2001 and 2008, when the system displayed reduction in the orbital period, before undergoing a fast brightening of several magnitudes \citep{tylenda2011v1309}. Before the outburst the object steadily brightened over 3\,yr. These time-scales and the concomitant characteristics of the contact-binary periodic variability are not easily explained with a common envelope dynamical phase, implying a slower interaction leading up to fast dynamical common envelope merger.

In order to explain the slower, pre-outburst rise of V1309~Sco, \citet{pejcha2016binary, pejcha2017pre} and \citet{metzger2017shock} took a different approach to simulating common envelope transients. Their study targeted the light curve of a contact binary before inspiral. By simulating the flow through the $L_2$ point, they claimed to have identified the cause of the slow pre-outburst light increase in V1309~Sco, as well as of other characteristics of transient lightcurves. Although they do not comment on the effects of $L_2$ mass-loss on the subsequent phase of dynamical inspiral, the implication is that the two phases are intimately connected.

\citet{macleod2018runaway, macleod2018bound} presented additional simulations of the Roche lobe overflow phase leading up to a common envelope inspiral. They simulated a 1\,\msun\ primary and a 0.3\,\msun\ companion, at the start of Roche lobe overflow (20.6\,\rsun, for their system). The simulation was not continued past the beginning of the dynamic inspiral. They showed the decay of the orbit to be driven by exchange of angular momentum from the companion to the mass flow, which is then lost from the system through the $L_2$ point. \citet{macleod2018bound} claim that, contrary to what was concluded by \citet{pejcha2017pre}, mass lost through the $L_2$ point is not unbound until much later in the simulations, when the companion is almost in contact with the primary. This bound outflow forms a large disk of material in the orbital plane which collimates the late time unbinding, forming a bipolar gas distribution.

In summary, much is yet to be learned about the common envelope interaction itself and the transient phenomena that it precedes. Including an extended phase of mass transfer and following the distribution of ejecta after the dynamic inspiral moves us towards an integrated understanding of these adjacent phases.

In Section~\ref{sec:simulationsetup} we summarise the setup for our simulations. In Section~\ref{sec:numerical}, we give an overview of the simulation behaviour and discuss the effect of increasing resolution on our simulations, along with the conservation of energy and angular momentum. In Section~\ref{sec:orbitalevolution}, we go over the pre-inspiral phase of our simulations, analysing Roche lobe overflow numerically with comparisons to analytical equations. Further, we discuss the different processes contributing to the the reduction in orbital separation both before and during the dynamic inspiral. We also discuss the creation of a decretion disk, and the presence of fallback material. Section~\ref{sec:planetarynebulae} contains a discussion of the formation of planetary nebulae from common envelopes, and the applications of our final gas distributions to further simulations. We conclude and summarise in Section~\ref{sec:conclusions}.

\section{Initial conditions}
\label{sec:simulationsetup}

The simulations presented here follow on from work done by \citet{passy2012simulating} and \citet{iaconi2017effect}. The simulations we analyse include one that was already presented by \citet{iaconi2017effect}, who list further details. In summary, we modelled a binary system made of a 0.88\,\msun, 90\,\rsun\ RGB primary with a 0.392\,\msun\ point mass core in orbit with a 0.6\,\msun\ point mass companion. We used three different resolutions of $8 \times 10^4$, $2.3 \times 10^5$ and $1.1 \times 10^6$ SPH particles. In addition to these simulations, a $2.3 \times 10^5$ particle simulation was also evolved with a primary star in corotation with the orbit. The resolution length in SPH, the smoothing length, is related to the local number density of particles. Initially, SPH particles in our high resolution simulation have smoothing lengths ranging between 0.2 and 9\,\rsun. For more details on the smoothing lengths in the different simulations, refer to Table~\ref{table:resolution}.

The core of our giant star and the companion are represented by point masses. These particles do not have associated internal energies or pressures, instead interacting purely gravitationally with other particles in the simulation. Their gravitational potentials have been softened by a cubic spline kernel, with a softening length of 3\,\rsun. This means that within a radius of 3\,\rsun, the potential of the point mass particles is almost flat \citep[for more information on the exact process, see][]{price2017phantom}, and outside a radius of twice the softening length, the potential is exactly the analytical form. Softened potentials are typically used when point mass particles are not allowed to accrete gas, which is the case in our simulations.

Before the binary simulations are carried out, the primary star is relaxed in the computational domain, as explained by \citet{iaconi2017effect}. This tends to result in a stellar structure with a slightly larger radius. The relaxed radii differ slightly with resolution, being 87, 91 and 93\,\rsun\ for the low, medium and high resolutions, respectively. We also note that we use the volume-equivalent radius (defined by \citealt{nandez2014v1309} to be the radius of a sphere with the total volume occupied by particles) for gas with a higher density than the initial surface density of the star. The corotating simulation starts with a primary star stabilised in the inertial frame. Once the simulation starts, not only is the star suddenly immersed in the companion's potential, but it is also spinning. The combination of the distortion and expansion resulting from these forces may lead to a larger mass transfer rate. We discuss this further in Section~\ref{sssec:masstransfer}. We were unable to perform all the simulations in the corotating frame, because the length of the Roche lobe overflow phase is approximately doubled, and so the simulations would have become prohibitively long.

We chose the initial orbital separation such that the Roche lobe of the giant is approximately equal to its stellar radius. With our parameters, the initial orbital separation is about 218\,\rsun\ \citep[by Eq.~2 of][]{eggleton1983approximations}. With this setup, a phase of mass transfer begins immediately upon starting the simulation.  

The reduction in separation for systems that are started well outside of Roche lobe contact is, in nature, driven by tidal interactions from distortions of the primary envelope by the potential of the secondary. This is also true in simulations, though the orbit decays more quickly as a result of oscillations artificially set in motion by the initial conditions. This said, starting the simulation with an orbital separation larger than 218\,\rsun\  results in a far more stable orbit and it requires longer simulation times to get to a state of mass transfer.

\section{Simulation overview}
\label{sec:numerical}

\begin{figure*}
	\includegraphics[width=\linewidth]{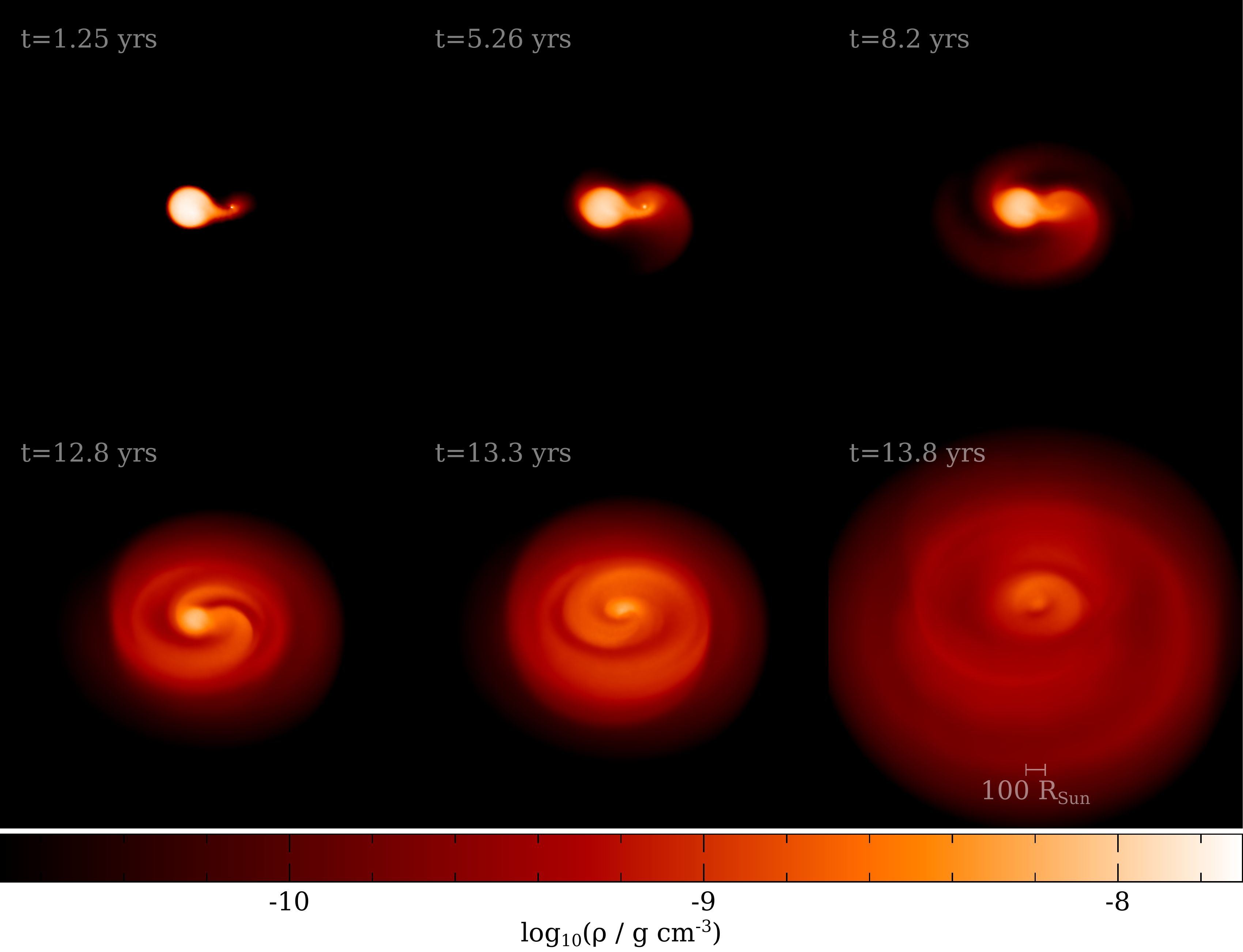}
    \caption{Surface rendering of our 1.1 million particle simulation. The frames are 5\,AU per side. From left to right, top to bottom, each row is a snapshot taken at 1.25, 5.26, 8.2, 12.8, 13.3 and 13.8\,yr. This is intended to display the overall evolution of the simulation. This image, along with all other renderings of the simulation in this paper, was created with \textsc{splash} \citet{price2007splash}. In surface renderings such as this, a value for the opacity is assigned arbitrarily for the desired visual result. In this case $\kappa = 1.9\times10^{-9}$\,cm$^2$\,g$^{-1}$.}
    \label{fig:evolution}
\end{figure*}

We first present an overview of the simulation parameters and their dependence on resolution, as well as conservation properties. We use this as a basis to describe the orbital evolution in Section~\ref{sec:orbitalevolution}. We show a visual overview of the simulation behaviour in Fig.~\ref{fig:evolution}. In the top three panels, the system is undergoing Roche lobe overflow. The bottom three panels show the evolution of the system within a year from beginning the dynamic inspiral.

\subsection{System evolution}
\label{ssec:overview}

\begin{figure*}
	\includegraphics[width=\columnwidth]{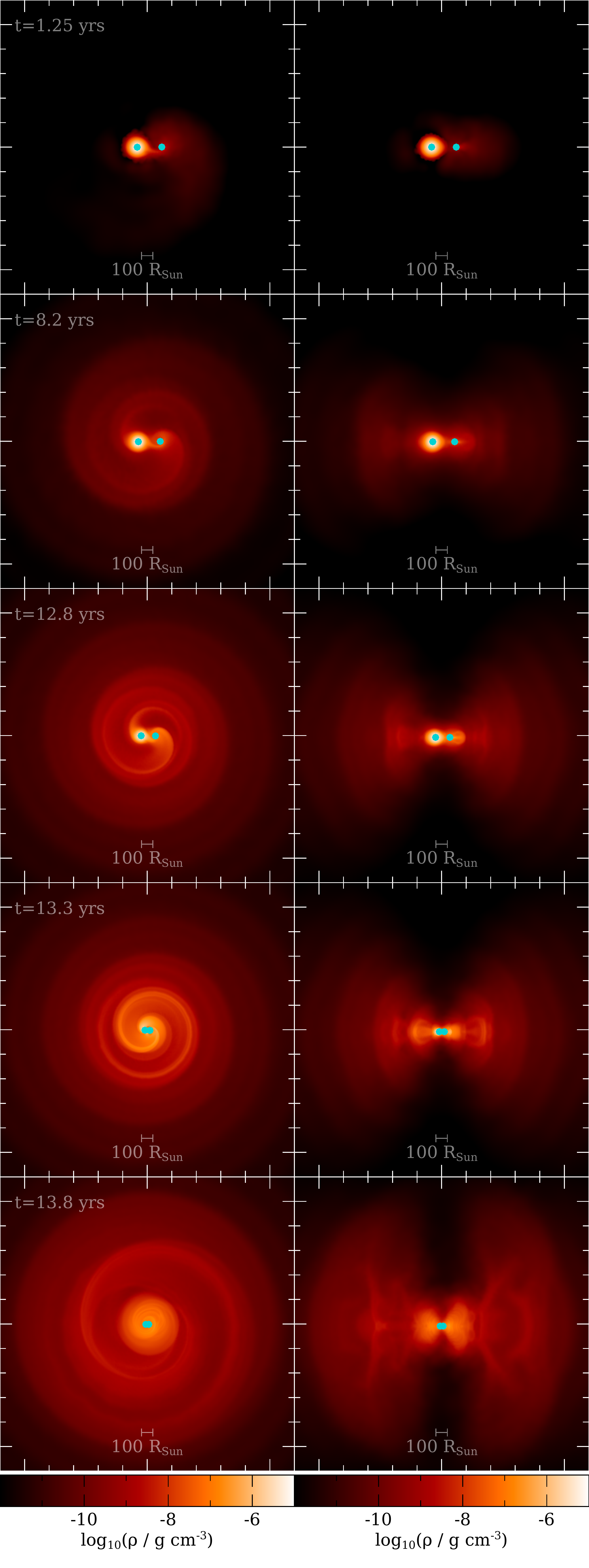}
	\includegraphics[width=\columnwidth]{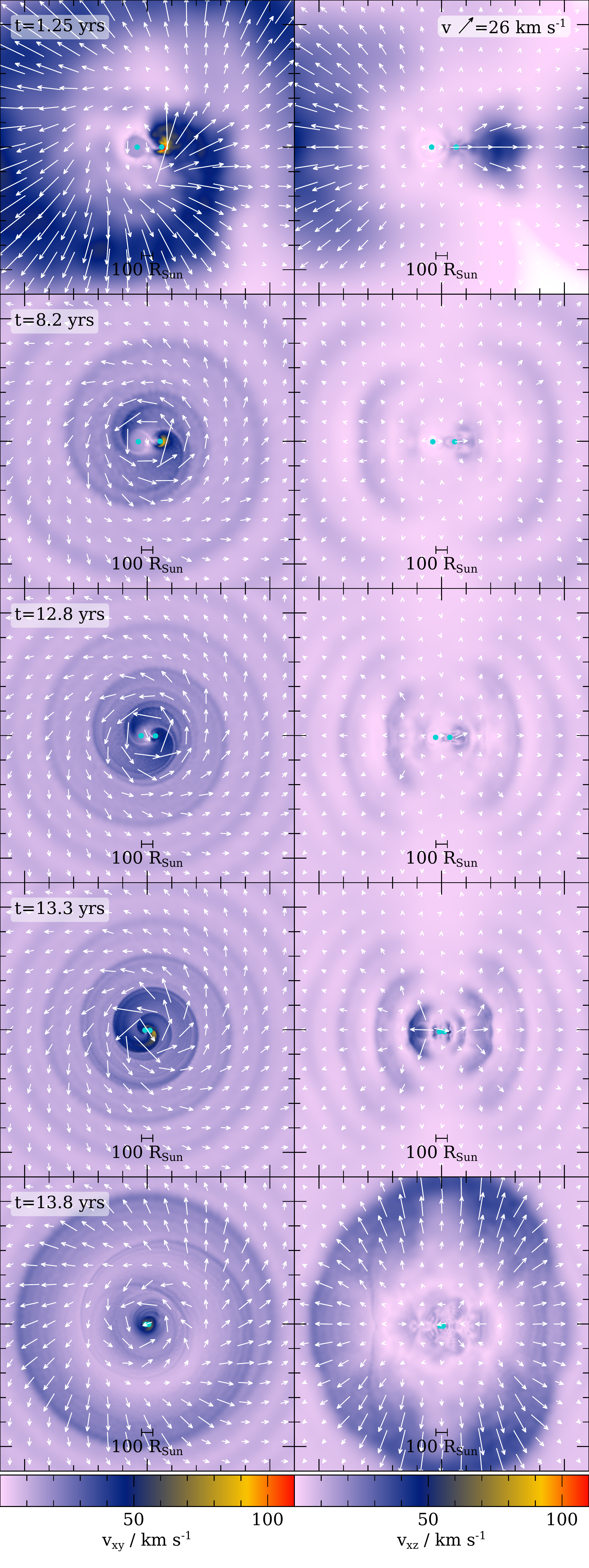}
    \caption{Columns 1 and 3: density and velocity slices in the orbital plane respectively. Columns 2 and 4: density and velocity slices in the perpendicular plane respectively. Arrows show the direction and relative magnitude of the velocity. Each of the frames is 12\,AU on each side and was taken from our 1.1 million particle simulation. The two point mass particles are plotted as green points, with the primary to the left and companion to the right. From top to bottom, each row is a snapshot taken at 1.25, 8.2, 12.8, 13.3 and 13.8\,yr.}
    \label{fig:slices}
\end{figure*}

The two left-hand columns of Fig.~\ref{fig:slices} shows the density evolution in the orbital plane (left) and along the plane perpendicular to the orbital plane (right) as a function of time (top to bottom) for the 1.1 million particle simulation. The two right-hand columns of Fig.~\ref{fig:slices} shows the corresponding velocity slices. At the start of the simulation, gas moves in a stream from the primary to the companion. This stream does not flow exactly through the inner Lagrange point $L_1$, instead flowing slightly around it. This can be explained by the star not being in corotation with the orbit, which is further evidenced by the fact that the stream does flow through the $L_1$ point in the corotating simulation. Some of this mass collects in a small disk around the companion, while the orbital separation decreases and the mass transfer rate increases. Eventually, gas starts flowing out of the system via the outer Lagrange point $L_2$ and eventually through $L_3$. These outflows are visible in Fig.~\ref{fig:slices} as spiral tails extending from both point masses in the orbital plane. The $L_2$ outflow is present to the right of the companion (right-hand green dot) in the top four rows. The $L_3$ outflow is most prominent in the second and third rows of Fig.~\ref{fig:slices} as a tail extending from the left of the primary.

Later, the companion is engulfed in the atmosphere of the donor star, and common envelope inspiral ensues on a dynamical time-scale, leading to a rapid reduction in the orbital separation (bottom row of Fig.~\ref{fig:evolution} and last two rows in Fig.~\ref{fig:slices}). As the companion is engulfed, a shell of material is ejected above escape velocity, becoming unbound. More material is unbound as the inspiral continues, but most of the envelope is only lifted to larger radii, remaining bound. This is consistent with the findings of \citet{iaconi2018effect}, that most of the unbound ejecta is from the surface of the donor star and is lost at the beginning of the dynamic inspiral. The mass that is ejected collides with material that previously flowed out through the $L_2$ and $L_3$ points, shaping the common envelope (we discuss this further in Section~\ref{ssec:gasdistribution}).

Due to evacuation of gas within the orbit of the far more compact binary, and a reduced velocity contrast between the cores and the gas due to the envelope being dragged into corotation, the inspiral slows down dramatically. There is still some interaction, hence the cores do not fully halt in their radial migration. Further, during this extended phase of slow inspiral, we begin to witness a partial fallback of some of the gas that was not fully unbound (see Section~\ref{ssec:fallback}). 

\subsection{Resolution Study}
\label{ssec:resolution}

\begin{table}
\begin{tabular}{x{1.1cm} x{1cm} x{1cm} x{1cm} x{1cm} x{1cm}}
\hline
$n_\text{part}$ & $h_\text{min;0}$ & $h_\text{max;0}$ & $h_\text{min;ins}$ & $h_\text{90;ins}$ & $h_\text{max;ins}$\\
 & / \rsun & / \rsun & / \rsun & / \rsun & / \rsun\\
\hline
$7.6 \times 10^{4}$ & 0.52 & 14 & 0.43 & 170 & $1.0 \times 10^4$ \\
$2.3 \times 10^{5}$ & 0.35 & 11 & 0.35 & 75 & $1.2 \times 10^4$ \\
$1.1 \times 10^{6}$ & 0.19 & 8.7 & 0.19 & 34 & $1.9 \times 10^4$ \\
$2.3 \times 10^{5}$\textsuperscript{\textdagger} & 0.35 & 11 & 0.26 & 390 & $2.0 \times 10^4$ \\
\hline
\multicolumn{6}{l}{\textsuperscript{\textdagger}\footnotesize{Simulation with corotating primary.}}
\end{tabular}
\caption{Comparison of SPH particle smoothing lengths, denoted here as $h$, for each of our different resolution simulations. The subscripts min and max refer to the outer limits of the range of $h$, while subscript 90 denotes the $h$ below which 90 per cent of particles lie. Subscript {\it 0} refers to the beginning of the simulation, while subscript {\it ins} refers to the time of fastest descent during the dynamic inspiral (6.4, 8.7, 13.3 and 18.0\,y for the low, medium and high resolution and corotating simulations, respectively).}
\label{table:resolution}
\end{table}

To see whether our results are sensitive to numerical resolution, we performed a resolution study. Because the highest resolution simulation, at 1.1 million SPH particles, required approximately 10 months on a 32 core server, increasing the resolution further is currently unfeasible. Instead, we compared our simulation with a series of lower resolution calculations. Table~\ref{table:resolution} lists the particle numbers and the corresponding minimum and maximum resolution lengths (the SPH smoothing length) in each simulation, both at the beginning of the simulation (second and third columns) and during the dynamical inspiral (fourth, fifth and sixth columns). It should be noted that the smoothing length of a particle scales inversely and smoothly with density. Hence the simulations are most well resolved around the cores, where the gas is most dense. We observe that the 1.1 million particle simulation has smoothing lengths typically smaller than half that of the 80\,000 particle simulation, except for the outer portions of the initial star.

\begin{figure}
	\includegraphics[width=\columnwidth]{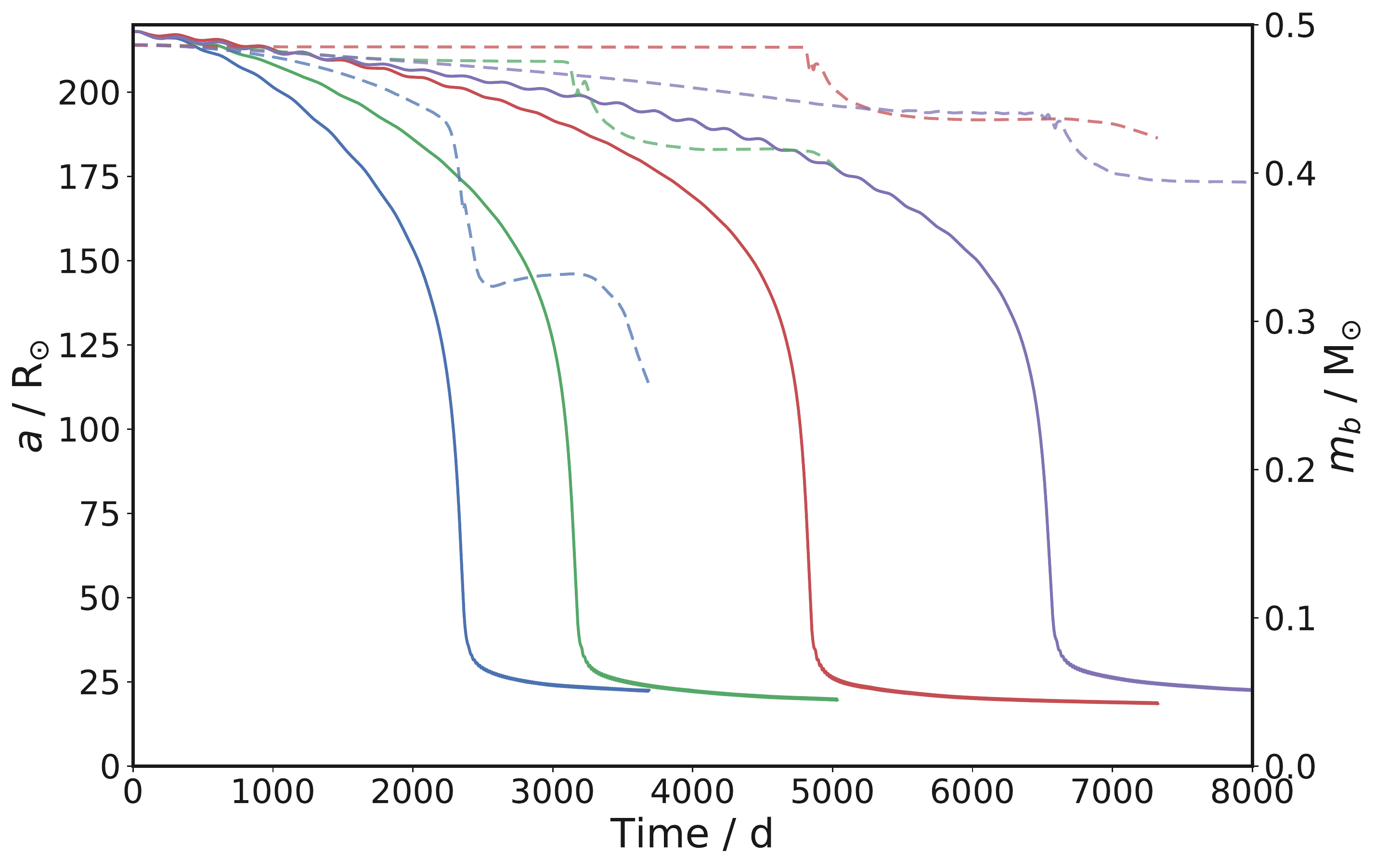}
    \caption{Comparison of the orbital evolution (solid lines) and bound mass (dashed lines) for simulations with differing resolutions. Resolutions are $8\times10^4$ particles (blue), $2.3\times10^5$ particles (green) and $1.1\times10^6$ particles (red). The $2.3\times10^5$ particle simulation was also run with a corotating primary (purple).}
    \label{fig:218sepbound}
\end{figure}

A comparison of the three simulations shows that the orbital separation at the end of the fast inspiral and the average rate of descent during the fast inspiral appear converged with respect to numerical resolution (see Table~\ref{table:convergence}). These data were collected from the simulation outputs by defining the dynamic inspiral to start and finish at times such that $|\frac{\dot{a}}{a}| \geq \frac{1}{15} \max |\frac{\dot{a}}{a}|$ holds true. For more discussion of the time-scales, see Section~\ref{sssec:orbitalevolutiontime-scale}. The full orbital evolution of the simulations is shown as the solid curves in Fig.~\ref{fig:218sepbound}. Table~\ref{table:convergence} also gives the amount of unbound material in our simulations, which is resolution dependent. In this work, we use the sum of kinetic and potential energies to determine unbound particles (if $E_\text{kin} + E_\text{pot} > 0$, the gas is unbound from the system). This gives a lower limit to the unbound mass. In some works the internal (thermal) energy is also considered ($E_\text{kin} + E_\text{pot} + E_\text{int} > 0$), which should lead to a greater amount of unbound gas. We include the amounts of unbound mass calculated with both criteria in Table~\ref{table:convergence}, but it is our experience that even including internal energy in the unbound criterion, the amount of unbound mass does not increase significantly \citep{staff2016agb}.

Fig.~\ref{fig:218sepbound} compares the mass of bound material for our three resolutions (dashed lines). Higher resolution simulations unbind less material. Gas is primarily unbound during the fast inspiral, although a small amount is unbound before, during Roche lobe overflow. This pre-inspiral unbinding is smaller at higher resolution, with the high resolution simulation unbinding only 0.05\,\msun\ before the inspiral. This is a consequence of our choice of initial conditions. Comparing the inertial and corotating frame simulations at intermediate resolution we see that the pre-inspiral unbinding is greater for rotating stars. The length of the pre-inspiral phase does not appear to have converged at the resolutions we have tested, meaning that it could be longer in reality.

Further, each of the simulations begins to experience renewed unbinding shortly after the time of orbital stabilisation. For the high resolution simulation, this begins to take place shortly before we stop the simulation (hence is not seen in Fig.~\ref{fig:218sepbound}), much longer after stabilisation than for the lower resolution simulation. This unbinding event is somewhat deceptive, in that it does not represent a large shift in the energy of particles that become unbound. In other words, particles that are close to the threshold of being unbound receive relatively small boosts in kinetic energy, from the orbital energy of the two cores, pushing the particles just over the threshold. Although the exact cause is unclear, this resolution-dependent unbinding occurs approximately when the softening radius of the point masses is no longer resolved in the simulation, which happens when the local smoothing length of the gas particles exceeds the softening radius. This occurs at late times because of the decreasing gas density surrounding the remnant cores.

\begin{table*}
\begin{tabular}{x{1.2cm} x{1.2cm} x{1.2cm} x{1.2cm} x{1.2cm} x{1.2cm} x{1.2cm} x{1.2cm} x{1.2cm} x{1.2cm} x{1.2cm}}
\hline
$n_\text{part}$ & $a_\text{i}$ & $a_\text{f}$ & $t_\text{f}$ - $t_\text{i}$ & $\frac{\Delta a}{\Delta t}$ & $m_\text{u,i}$ & $m_\text{u,p}$ & $\frac{m_\text{u,p}}{m_\text{tot}}$ & $m_\text{u,i}$\textsuperscript{*} & $m_\text{u,p}$\textsuperscript{*} & $\frac{m_\text{u,p}\textsuperscript{*}}{m_\text{tot}}$\\
 & / \rsun & / \rsun & / d & / \rsun\,d$^{-1}$ & / \msun & / \msun & $\times 100$ & / \msun & / \msun & $\times 100$\\
\hline
$7.6 \times 10^{4}$ & 130 & 30 & 276 & 0.36 & 0.045 & 0.157 & 32.2 & 0.053 & 0.160 & 32.9\\
$2.3 \times 10^{5}$ & 126 & 28 & 275 & 0.36 & 0.011 & 0.071 & 14.6 & 0.013 & 0.076 & 15.6\\
$1.1 \times 10^{6}$ & 122 & 28 & 263 & 0.36 & 0.001 & 0.050 & 10.3 & 0.002 & 0.054 & 11.1\\
$2.3 \times 10^{5}$\textsuperscript{\textdagger} & 129 & 31 & 354 & 0.28 & 0.046 & 0.092 & 18.9 & 0.048 & 0.097 & 19.9\\
\hline
\multicolumn{7}{l}{\textsuperscript{\textdagger}\footnotesize{Simulation run with corotating primary.}}\\
\multicolumn{7}{l}{\textsuperscript{*}\footnotesize{Unbound criterion includes internal energy.}}
\end{tabular}
\caption{Comparison of orbital evolution quantities for different resolutions. Here we choose $a$ as the orbital separation, $m_\text{u}$ as the unbound mass, $m_\text{tot} = 0.49$\,\msun\ as the total gas mass in the simulation and $t$ as the time. The subscripts i and f refer to times at the beginning and end of the fast inspiral as defined in the text, respectively. The subscript p refers to the `plateau' value of the unbound mass, after the end of the fast inspiral. $\frac{\Delta a}{\Delta t}$ is the mean rate of descent during the fast inspiral.}
\label{table:convergence}
\end{table*}

\subsection{Energy and Angular Momentum Conservation}
\label{ssec:conservation}

\begin{figure}
    \includegraphics[width=\columnwidth]{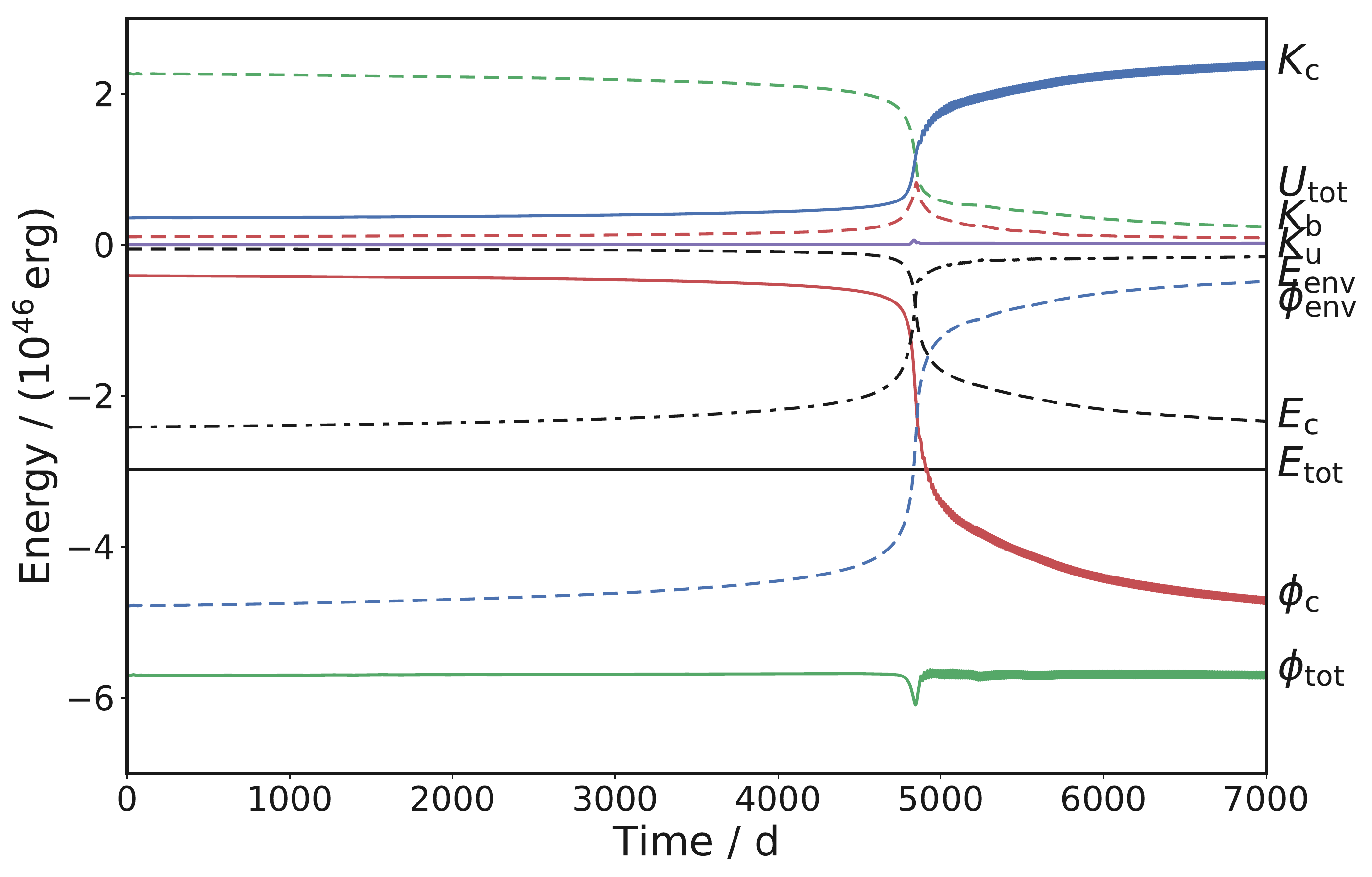}
    \caption{Evolution of the energy components in time. Here $K_\text{c}$ (solid blue) is the total kinetic energy of the two cores, $U_\text{tot}$ (dashed green) is the total internal energy of the gas, $K_\text{b}$ (dashed red) is the kinetic energy of bound gas, $K_\text{u}$ (purple) is the unbound kinetic energy, $E_\text{env}$ (dot-dashed black) is the envelope energy, $\phi_\text{env}$ (dashed blue) is the potential energy of the envelope, $E_\text{c}$ (dashed black) is the orbital energy of the cores, E$_\text{tot}$ (solid black) is the total energy of the system, $\phi_\text{c}$ (solid red) is the potential energy between the point mass particles and $\phi_\text{tot}$ (solid green) is the total potential energy.}
    \label{fig:1e6energy}
\end{figure}

\begin{figure}
	\includegraphics[width=\columnwidth]{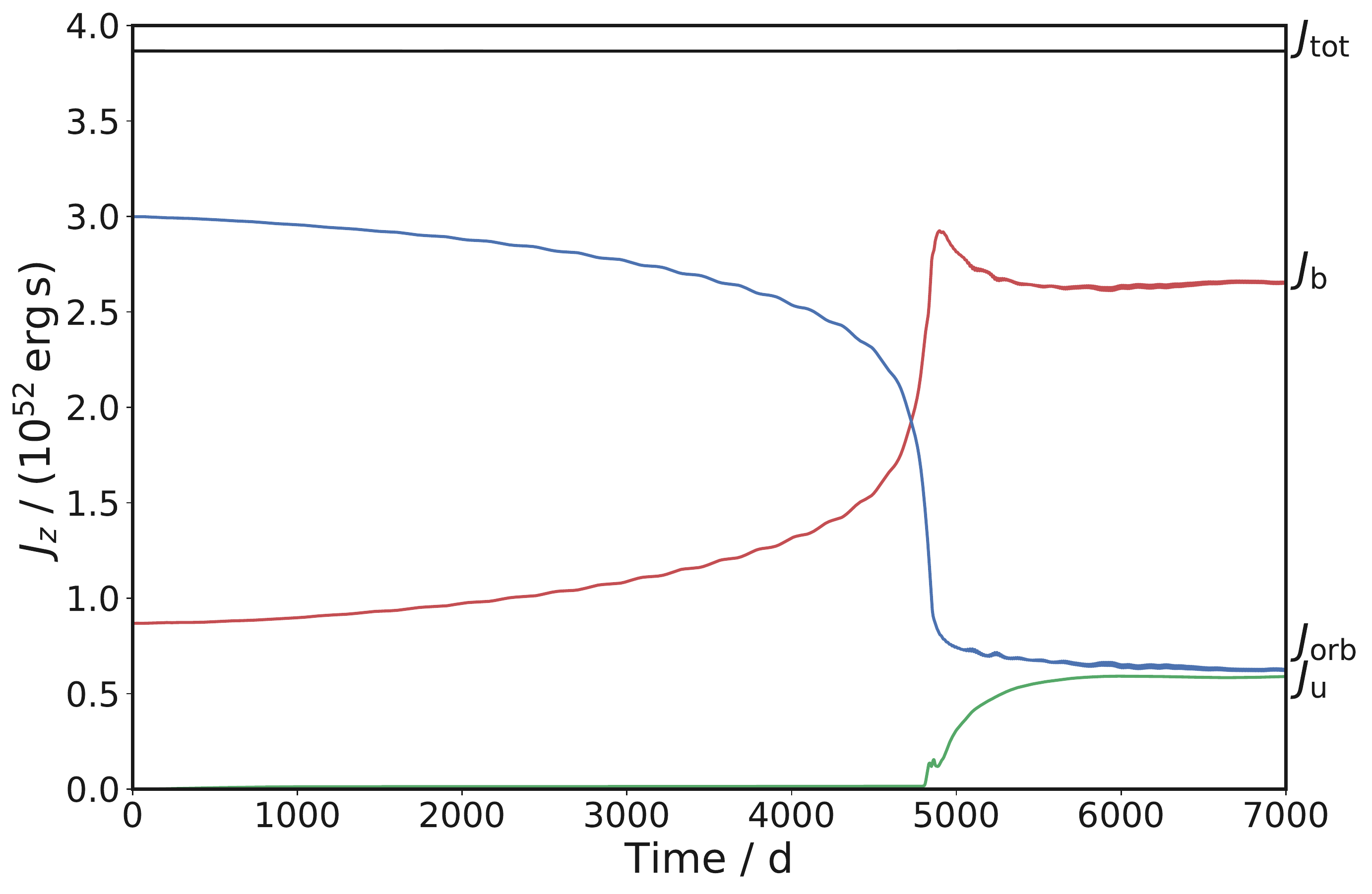}
    \caption{Evolution of the angular momentum components in time. In this plot: $J_\text{b}$ (red line) is the angular momentum of bound gas, $J_\text{orb}$ (blue line) is the angular momentum of the point mass particles, $J_\text{u}$ (green line) is the angular momentum of unbound material, and $J_\text{tot}$ (black line) is the total angular momentum in the simulation.}
    \label{fig:1e6angmom}
\end{figure}

Our simulations employed a single, global timestep for all particles, which leads to excellent conservation of energy and angular momentum, at the cost of computational speed. Both physical quantities are conserved to better than 0.1 per cent, an improvement on prior SPH simulations \citep[e.g. the 1 per cent conservation observed in the simulations of ][]{passy2012simulating}. Figs~\ref{fig:1e6energy} and \ref{fig:1e6angmom} show the components of energy and angular momentum in our highest resolution simulation. Conservation is approximately 0.1 per cent and 0.3 per cent for our medium and low resolution simulations, respectively.

In Figs~\ref{fig:1e6energy} and \ref{fig:1e6angmom}, we plot multiple quantities as a direct comparison to the equivalent figures in \citet{passy2012simulating}. For much of the simulation, none of the components vary significantly. Our primary core and companion particles do not possess internal energy because they interact with the SPH particles only gravitationally. Hence the total internal energy is summed over only the gas particles. All other quantities are found by summing components from gas and point mass particles as needed. Of note is that the envelope energy, calculated as the sum of the envelope potential energy, the total internal energy and the bound kinetic energy, begins negative and remains so for the entire simulation. The fact that by the end of the simulation the average total energy of the envelope is close to zero suggests that the envelope is only marginally bound. It is possible that with the help of part of the recombination energy budget this may lead to complete unbinding of the envelope.

Fig.~\ref{fig:1e6angmom} shows only the z-components of the angular momentum (with respect to the centre of mass of the system at the beginning of the simulation, located at the origin, because the majority of rotational motion is in the orbital plane of the system). We plot the angular momentum of the bound and unbound gas, as well as the angular momentum of the two cores. There is a slight peak in the angular momentum curve of the bound mass. This is because, after this point, some mass becomes unbound. The excellent conservation gives an indication that the transfer of angular momentum is being handled correctly within our simulation. The analytical equation for the angular momentum of a binary system is

\begin{equation}
J = M_1 M_2 \left(\frac{G a}{M}\right)^{\frac{1}{2}},
\label{eq:j}
\end{equation}
where $M_1$, $M_2$, $M$, $a$ and $G$ are the primary and secondary star masses, the total mass of the system, the orbital separation and the gravitational constant, respectively. Using this, we calculate a total angular momentum budget of $3.87 \times 10^{52}$\,erg\,s. This is consistent with that calculated directly from the simulation.

\subsection{The Corotating Reference Frame}
\label{ssec:corotating}

One of our two simulations with an intermediate resolution ($2.3 \times 10^{5}$ particles) was carried out in a frame that was rotating with the same angular velocity as the initial binary system. The primary star was stabilised in isolation in the inertial reference frame. Then it and the companion were placed in the corotating frame, with the gas particles set to be initially stationary in this frame. Ideally, the giant star should be stabilised alongside its companion in the corotating frame, and this step will be included in future simulations.

The primary difference between use of a non-rotating primary star and one which is tidally locked with its companion is that the tidally locked star overfills its Roche lobe more. This would suggest that the initial mass transfer should be larger than for simulations with a non-rotating primary.

This simulation progressed over a time-scale approximately double that of the equivalent inertial frame simulation. This is because gas particles begin the simulation with greater angular momentum, hence a gas particle requires less additional angular momentum to be ejected. Point mass particles then exchange less angular momentum with the gas as it escapes, resulting in a less rapid decrease in orbital separation. 

As the orbital separation decreases in the next phase, the relative velocity between the companion and the gas increases, but due to the initial rotation of the primary, the gas is dragged into corotation with the cores more rapidly than for the inertial frame simulation. Hence, the orbit enters the self-regulated inspiral at 31\,\rsun, a slightly larger separation than 28\,\rsun, measured from the equivalent, non-rotating simulation. The mean rate of inspiral is also slower in the corotating simulation (compare 0.28 \rsun\,d$^{-1}$ and 0.36 \rsun\,d$^{-1}$ in Table~\ref{table:convergence}).

\section{Orbital evolution}
\label{sec:orbitalevolution}

Having discussed the numerical aspects of the simulation, we here describe in detail the phases of our simulation that were outlined in Section~\ref{ssec:overview}.

\subsection{Roche lobe Overflow}

At the onset of Roche lobe overflow, we expect the systems to conserve angular momentum, and for the mass transfer rate to be related to the amount by which the primary overfills its Roche lobe.

\subsubsection{Orbital evolution time-scale}
\label{sssec:orbitalevolutiontime-scale}

The primary star is not stabilised in the potential of the companion, so the giant is not in perfect equilibrium at the start of the simulation. The distortions to the primary envelope that result are small, but likely lead to an orbital evolution that is too fast. In addition, the rate at which the Roche lobe overflow phase proceeds and how quickly the system enters a common envelope, depends on resolution (see Fig.~\ref{fig:218sepbound}). At our lowest resolution of approximately 80\,000 SPH particles, the Roche lobe overflow phase progresses in about half the time required for our high resolution simulation of 1.1 million particles. This by itself demonstrates that we cannot trust the simulated length of the Roche lobe overflow phase. This reaction to the new potential is also likely to be the cause of the initial mild eccentricity, visible in Fig.~\ref{fig:218sepbound}, that develops in the orbit early in the simulation. 

Mass transfer between the stars and mass loss from the $L_2$ and $L_3$ locations drives orbital shrinkage (see Eq.~\ref{eq:RLOF}), and it is likely affected by resolution, because the mass is discretised differently. Below we compare the inverse of the orbital time-scale, $\frac{\dot{a}}{a}$, calculated purely from the orbital separation of the two cores in the simulation with that derived from Eq.~\ref{eq:j} and Kepler's third law,

\begin{figure}
	\includegraphics[width=\columnwidth]{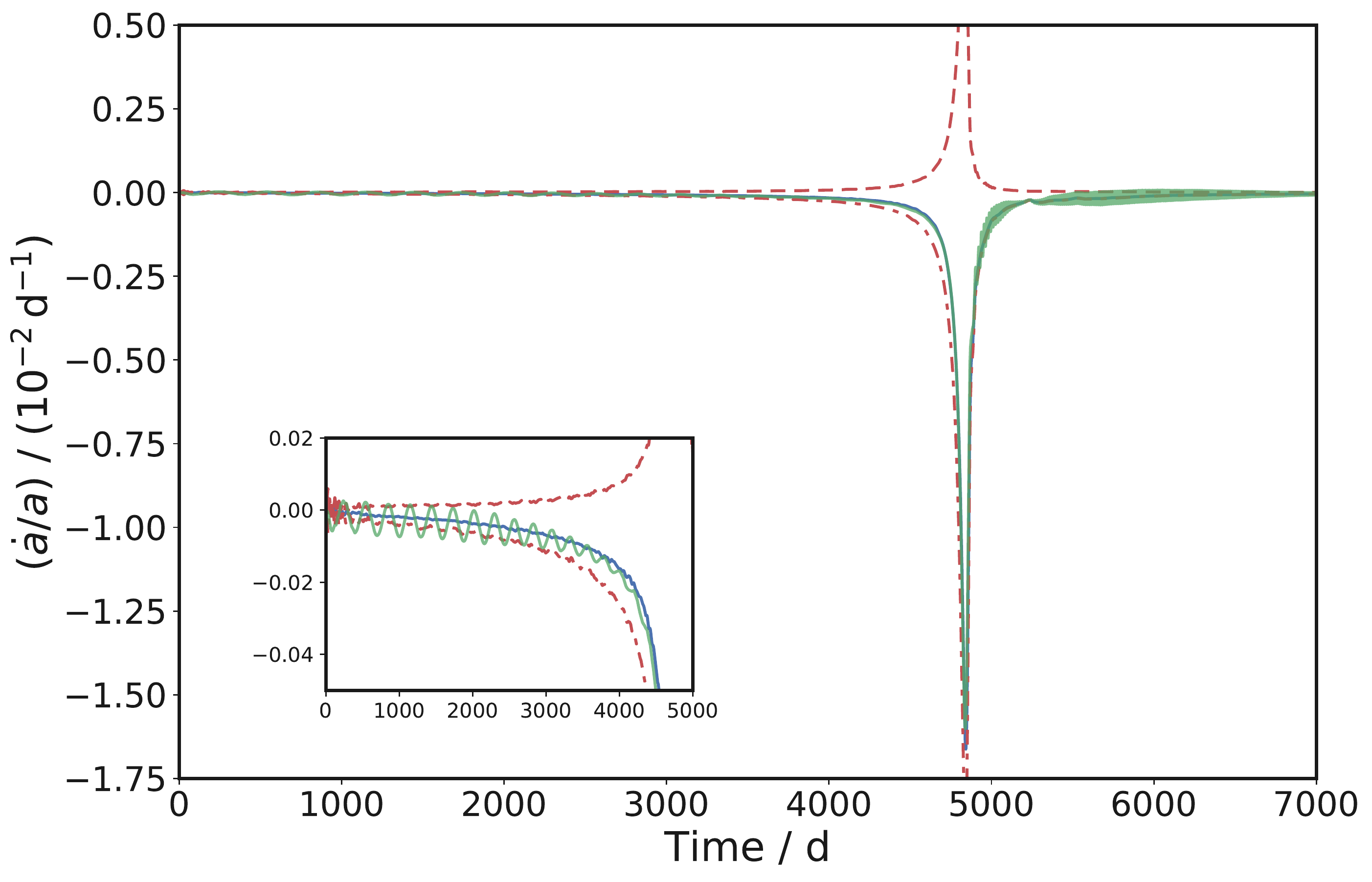}
    \caption{Comparison of the simulated orbital evolution timescale (solid green line) for the 1.1 million SPH particle simulation, with the analytic time-scale computed from Eq.~\ref{eq:RLOF} (solid blue line). Also displayed lines computed with Eq.~\ref{eq:RLOF} where we have set $\dot{J} = 0$ (dashed red line) and $\dot{M_2} = - \dot{M_1}$ (dash-dotted red line).}
    \label{fig:RLOF}
\end{figure}

\begin{equation}
\frac{\dot{a}}{a} = \frac{2 \dot{J}}{J} - \frac{2 \dot{M_1}}{M_1} - \frac{2 \dot{M_2}}{M_2} + \frac{\dot{M_1} + \dot{M_2}}{M_1 + M_2},
\label{eq:RLOF}
\end{equation}

\noindent where the symbols have the same meaning as in Eq.~\ref{eq:j}.

In Eq.~\ref{eq:RLOF}, we use the mass contained in the primary and companion Roche lobes ($M_1$ and $M_2$, respectively). For the angular momentum $J$, we sum the contributions from the two core particles and the angular momentum of the gas inside the Roche lobes of the binary, all with respect to the centre of mass. During the interaction, some gas is expelled from the Roche lobes of the binary, forming the spiral density arms that are visible in Fig.~\ref{fig:slices}. The angular momentum carried off in the spiral waves drives the decrease in orbital separation. It should be noted that the separation should decrease even if $\dot{J} = 0$, as long as $M_1 > M_2$ and $\dot{M_2} = - \dot{M_1}$. The disk formed by this outflow is discussed in Section~\ref{sssec:edisk}.

 We measure the mass transfer rate in the equation, $\dot{M_1}$, by counting the SPH particles that have entered the Roche lobe of the companion star for the first time, and then dividing this mass by the time between two simulation outputs. The Roche lobe radius is updated at every code output to reflect the changes in orbital separation and stellar masses. This method of measuring mass transfer is an approximation because, for example, it does not take into account particles that enter the companion's Roche lobe but exit it before the next code output. 

The simulated and analytically derived time-scales, computed for the highest resolution simulation, compare to within 10 per cent (Fig.~\ref{fig:RLOF}, green and blue lines, respectively), although the numerical time-scale displays periodic behaviour because of the slight orbital eccentricity. This means that the relationship between mass transfer rate, angular momentum loss and orbital separation are in line with the analytical prescription.

We further plot in Fig.~\ref{fig:RLOF}, Eq.~\ref{eq:RLOF}, but with $\dot{J} = 0$ (dashed red line), or with $\dot{M_2} = - \dot{M_1}$ (dash-dot red line). In the case where $\dot{J} = 0$, $\dot{a}/a$ is positive throughout, while if $\dot{M_2} = - \dot{M_1}$, $\dot{a}/a$ is mainly negative, and similar to the observed orbital decay. By separating the effect of mass transfer from the effect of angular momentum loss, we can deduce that it is primarily the loss of angular momentum via $L_2$ and $L_3$ that drives the decrease in orbital separation. 

After the end of the Roche lobe overflow phase, at around 4600\,d in the high resolution simulation, dynamic inspiral begins. At this time Eq.~\ref{eq:RLOF} can no longer be used to predict the time-scale accurately, because it becomes difficult to define what is meant by the masses of the stars and their ejecta when they are orbiting within a common envelope, so the time-scales do not match as closely as before the dynamic inspiral but still display the same basic shape. 

\subsubsection{The Mass Transfer Rate}
\label{sssec:masstransfer}

We next investigate the mass transfer during the Roche lobe overflow phase and the ability of simulations to reproduce it. The aim is to determine whether the simulations can indicate the duration of the unstable Roche lobe overflow phase, and ultimately lead to a better determination of the criterion of instability.

For the simple case of a binary with a non-spinning donor, which can be modelled as a polytrope, and for which the mass transfer is perfectly conservative, unstable mass transfer occurs for $q \gtrsim \frac{2}{3}$ or 0.67 \citep{tout1991wind}. The mass ratio of our system is almost exactly $q = 0.68$, our primary star starts with no spin, and is fairly well approximated by a polytrope of index $n = 1.5$ (suitable for red giant stars). Our mass transfer is not conservative, though the mass lost is relatively low, only about 7 per cent over the entire Roche lobe overflow phase. It is therefore expected that our system should be unstable, as demonstrated by our simulations.

\begin{figure}
	\includegraphics[width=\columnwidth]{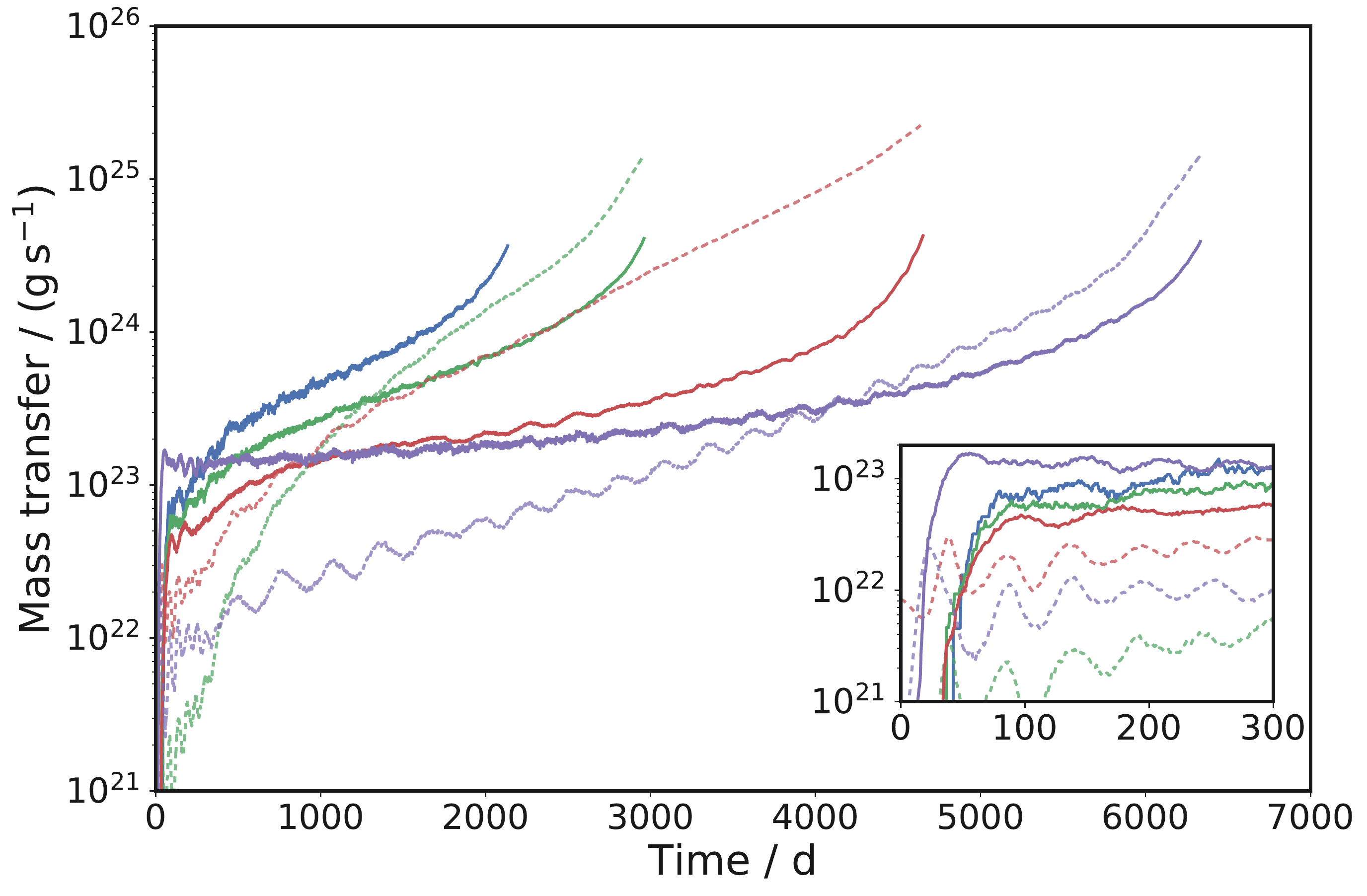}
    \caption{Mass transfer rate through $L_1$ as a function of time. As previously, blue, green and red lines represent the low, medium and high resolution simulations, respectively, while the purple line is used for the medium resolution simulation in a corotating frame. Main panel: solid lines are mass transfer rates measured from the simulations. Dashed lines represent the analytical form of the expected mass transfer rate given inputs from their respective simulations (see Eq.~\ref{eq:mdot}). Insert: same as the main panel, but for the first 300 days of the simulations.}
    \label{fig:massflow}
\end{figure}

In Fig.~\ref{fig:massflow} we plot the simulated mass transfer rate determined as described in Section~\ref{sssec:orbitalevolutiontime-scale}. The lines are drawn only to show the time span of interest, namely before the system enters the dynamic inspiral. As mentioned in Section~\ref{ssec:resolution}, this time is chosen as the first point when |$\frac{\dot{a}}{a}| \geq \frac{1}{15} \max |\frac{\dot{a}}{a}|$. After that time (about 4600\,d in the high resolution simulation) the Roche geometry is no longer applicable and measuring mass transfer from one star to another no longer makes sense. 

As expected, the transfer of mass from the primary to the companion starts slowly (about $10^{22}$\,g\,s$^{-1} = 6.6 \times 10^{-6}$\,\msun\,yr$^{-1}$), and increases by two orders of magnitude as the orbital separation decreases. Because of the discretisation of mass in SPH simulations, the minimum (non-zero) rate of mass transfer is when one SPH particle passes $L_1$ between two consecutive code dumps (dumps are appoximately every 80\,000 seconds). This corresponds to mass transfer rates of $10^{23}$\,g\,s$^{-1}$ and $10^{22}$\,g\,s$^{-1}$, for the lowest and highest resolutions we adopt, respectively. The measured mass transfer rate curves in Fig.~\ref{fig:massflow} have been smoothed with a Savitzky-Golay filter, showing the underlying trends in the mass transfer measurements (which can otherwise be noisy when close to the minimum rates), so the plotted rate can be smaller than the minimum simulated rate. It is important to stress that this is a result of smoothing the curves, and no rate below the minimum is actually measured. We can only consider the measured mass transfer rates after a simulation time of 300\,d, when the transfer rates, at all resolutions, are larger than the minimum measurable. 
At a time of 300\,d mass transfer rates are quite similar for all resolutions, varying from $10^{23}$\,g\,s$^{-1}$ at our lowest resolution to $5\times10^{22}$\,g\,s$^{-1}$ at the highest resolution.

A question is whether we can verify these mass transfer rates. An analytical expression for stable and conservative mass transfer rate though the $L_1$ point is that of J\k{e}drzejec \citep{paczynski1972evolution}, 

\begin{equation}
\dot{M}_1 = - S_1 \left(\frac{\mu m_H}{k_\mathrm{B} T}\right)^{1.5}W(M_\text{1,rel})\rho G^2 M_1^2 \left(\frac{\Delta R}{R_1}\right)^3,
\label{eq:mdot}
\end{equation}

\noindent where $S_1 \approx 0.215$ is a constant factor related to the polytropic index, $\mu$ is the mean molecular weight, $m_H$ is the mass of a hydrogen atom, $k_\mathrm{B}$ is Boltzmann's constant, $T$ is the temperature at the photosphere of the donor, $\rho$ is its density at the photosphere, $G$ is the gravitational constant, $M_1$ is the mass of the primary and $\Delta R = R_1 - R_\text{L,1}$, where $R_1$ is the radius of the primary and $R_\text{L,1}$ is its Roche lobe radius. The quantity $W(M_\text{1,rel})$ is a function of $M_\text{1,rel}$ = $M_1$/($M_1$ + $M_2$), defined as
\begin{equation}
W(M_\text{1,rel}) = \frac{\sqrt{M_\text{1,rel}}\sqrt{1-M_\text{1,rel}}}{(\sqrt{M_\text{1,rel}} + \sqrt{1-M_\text{1,rel}})^4}\left(\frac{a M_\text{1,rel}}{R_\text{L,1}}\right)^{n+1.5},
\end{equation}

\noindent where $a$ is the orbital separation and $n$ is the polytropic index, taken to be $1.5$ for a red giant. A similar derivation for $\dot{M}_1$ with an isothermal flow was made by \citet{ritter1988turning}, although we use Eq.~\ref{eq:mdot} because it is valid for adiabatic flows.

The mass transfer rate Eq.~\ref{eq:mdot} applies to giants transferring mass conservatively. It assumes a system with a corotating primary star and that, as the mass flows though the $L_1$ point, it transitions from subsonic to supersonic. The first of these conditions is only true for one of our simulations. However, all of the simulations do have a sonic point between the two stars. With this in mind, we apply it to our simulations, but acknowledge that there are some discrepancies resulting from the non-rotating primary.

To determine the mass transfer rate from Eq.~\ref{eq:mdot}, we used inputs from the initial stellar profile, specifically $\mu = 1.34$, $T = 3500$\,K and $\rho = 7\times 10^{-9}$\,g\,cm$^{-3}$. These were kept constant over time owing to difficulties in measuring them from the simulation, generating uncertainty. When the star is distorted by the presence of the companion, it is not trivial to determine its radius. This difficulty is compounded by material filling the orbit after having been stripped from the primary during Roche lobe overflow. Hence, in this case we simply determine the average radius of the stellar gas distribution whose surface has the approximate density of the stellar photosphere at the start of the simulation.

The analytical form of the mass transfer rate (Eq.~\ref{eq:mdot}) is compared with the measured numerical rate in Fig.~\ref{fig:massflow} for the high and intermediate resolution simulations (red and green lines) and for the simulation carried out in the corotating frame (purple line), but our discussion below is valid for the other simulations as well. 

The analytical mass transfer rate is slightly lower than the numerical rate at 300\,d, our fiducial start of the simulation as described above and the difference is smaller for higher resolutions. Both the simulation and the analytical approximation indicate that the mass transfer rate should increase in time, although the numerical rate starts larger but ends smaller than the analytical one. Finally, the analytical mass transfer rate is about four times larger than the numerical rate at the end of the higher resolution simulation and three times larger at the end of both the inertial and corotating frame, intermediate resolution simulations. Considering the imprecision with which we define the stellar radius and the third power dependence of the mass transfer rate on that number, the discrepancy here is relatively small. 

\subsubsection{The Decretion Disk}
\label{sssec:edisk}

\begin{figure*}
	\includegraphics[width=\linewidth]{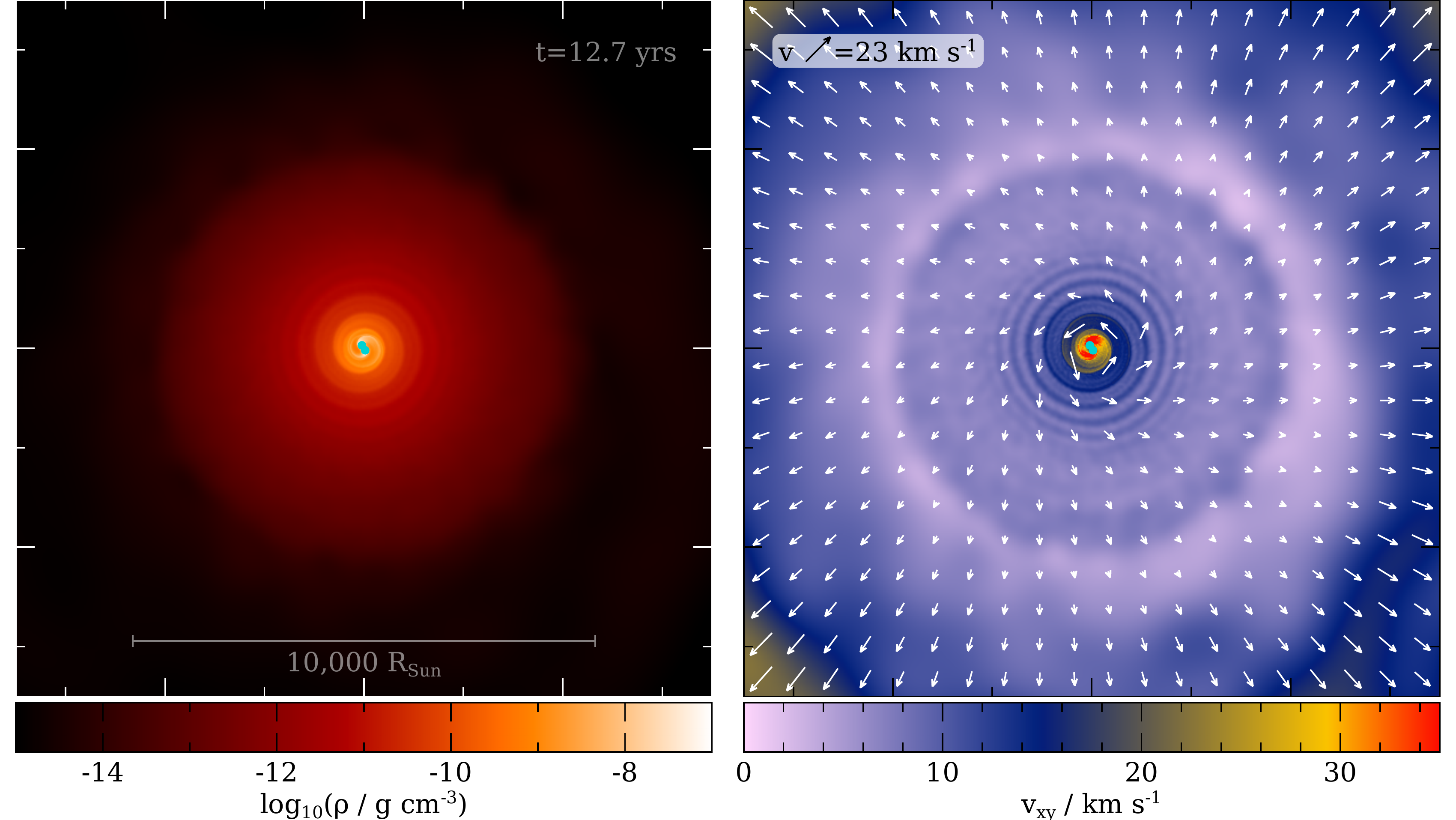}
    \caption{Cross sections of density (left) and velocity (right) in the orbital plane of the excretion disk at the beginning of the dynamic inspiral, at 12.8\,yr, for the 1.1 million particle simulation. These frames are taken at the same time as the third row of Fig.~\ref{fig:slices} (12.8\,yr), but are zoomed out such that the frames are 70\,AU per side. The green dots represent the cores.}
    \label{fig:disk}
\end{figure*}

As was mentioned in Section~\ref{sssec:orbitalevolutiontime-scale}, gas is ejected during Roche lobe overflow in the form of spiral density waves originating at $L_2$, and later at $L_3$. From our high resolution simulation we estimate that over the course of the Roche lobe overflow phase, these waves carry approximately $9 \times 10^{51}$\,erg\,s of angular momentum away from the binary (23 per cent of the total angular momentum budget of $3.87 \times 10^{52}$\,erg\,s) and about $7.8 \times 10^{-2}$\,\msun, driving the decrease in the orbital separation. The decretion disk is contained within approximately 5000\,\rsun\ from the centre of the binary (see Fig.~\ref{fig:disk}). Within the disk, the range of smoothing lengths $h$ is about 30 to 1000\,\rsun. Approximately $1.5 \times 10^{-3}$\,\msun\ of gas lies outside this radius and contains gas unbound from the system within the first year of the simulation.

\citet{tocknell2014constraints} performed a calculation (their Section~5), using data from the simulations of \citet{passy2012simulating}, estimating the final radius of a disk of material that would be formed by infalling bound gas at the end of the common envelope. Their estimate considered gas moving in ballistic trajectories, and used conservation of energy and angular momentum to determine the radius and time-scale of formation. Using the quantities in the previous paragraph, we calculate a specific angular momentum for the disk of $5.8 \times 10^{19}$\,erg\,s\,g$^{-1}$. Hence, following the procedure of \citet{tocknell2014constraints}, we find that this bound disk would fall back to a radius of about 130\,\rsun\ from the centre of mass. This is well outside the final orbital separation of the system. The time taken for this trajectory would be at least 20\,yr, though this assumes that the bound disk gas has no left over outward motion, making the timescale a lower limit. However, in this simulation this structure is unlikely to survive for long enough, because a dense, fast outburst occurs at the beginning of the dynamic inspiral and quickly overtakes the expanding disk. We discuss in Section~\ref{ssec:gasdistribution} the bipolar gas distribution at the end of our simulation.

\subsection{The Dynamical Inspiral Phase}
\label{ssec:inspiral}

At some stage in a common envelope interaction, the orbital separation starts to decrease at a faster rate. This occurs in our high resolution simulation at approximately 4600\,d, as indicated by the time-scale of the system (Fig.~\ref{fig:RLOF}). At this point, the mass transfer is no longer characterised by the mechanism of Roche lobe overflow because the system is no longer described accurately by the Roche geometry.

As the companion splashes into the atmosphere of the primary it unbinds a shell of material which travels outwards at speeds of about 40\,km\,s$^{-1}$. This is visible in Fig.~\ref{fig:slices} at 13.3\,yr, and later on as it has travelled outwards at 13.8\,yr, located at approximately 1000\,\rsun\ from the binary at the centre. We can also see in the perpendicular plane (far right column) of Fig.~\ref{fig:slices} at 13.8\,yr how the material that expands above and below the orbital plane is faster, while the material in the orbital plane encounters the torus and decelerates. This effect is similar to what is observed in the simulations of \citet{macleod2018bound}.

The binary has entered a common envelope and the inspiral is now driven by gravitational and hydrodynamic drag forces. Gravitational drag operates purely through gravitational interactions between the gas and the companion, while hydrodynamic drag is instead the force felt by a body owing to direct collisions between the body and the particles of the ambient medium through which it is moving. Hydrodynamic drag tends to be insignificant in typical binary common envelope interactions, while for lighter companions, such as planets, it can be relatively more important due to a lower intensity of the gravitational drag \citep{staff2016hydrodynamic}. We can only simulate gravitational drag, because the point mass particles have no physical size. This said, gas tends to collect in the shallow potential wells around point masses and travels with them, something that does effectively give them a radius. However, any collisions between the ambient medium and the material in the potential wells still only affects the cores via gravitational interactions. 

While one could expect that the strength of the drag is affected by resolution, it has been shown previously (e.g.,~\citealt{passy2012simulating,iaconi2017effect}) that it is only marginally so. This means that resolution does not dramatically affect the final separation or the speed of the inspiral (we can appreciate this by looking at Table~\ref{table:convergence}). \citet{staff2016hydrodynamic} showed that in the case of their interaction with a planetary mass companion, the strength of the gravitational drag was commensurate with that determined by an analytical approximation. Here we carry out a similar comparison for our stellar-mass companion. Below we compare the drag calculated in three distinct ways, two from the simulations directly and one using an analytical prescription together with simulation quantities, as done before.

The physical parameters that influence the gravitational drag are the mass of the embedded body, the density of the surrounding fluid and the velocity contrast of the body and the fluid. In a common envelope, however, there may be additional factors playing a role, such as the density gradient \citep{macleod2017common}.

\begin{figure}
	\includegraphics[width=\columnwidth]{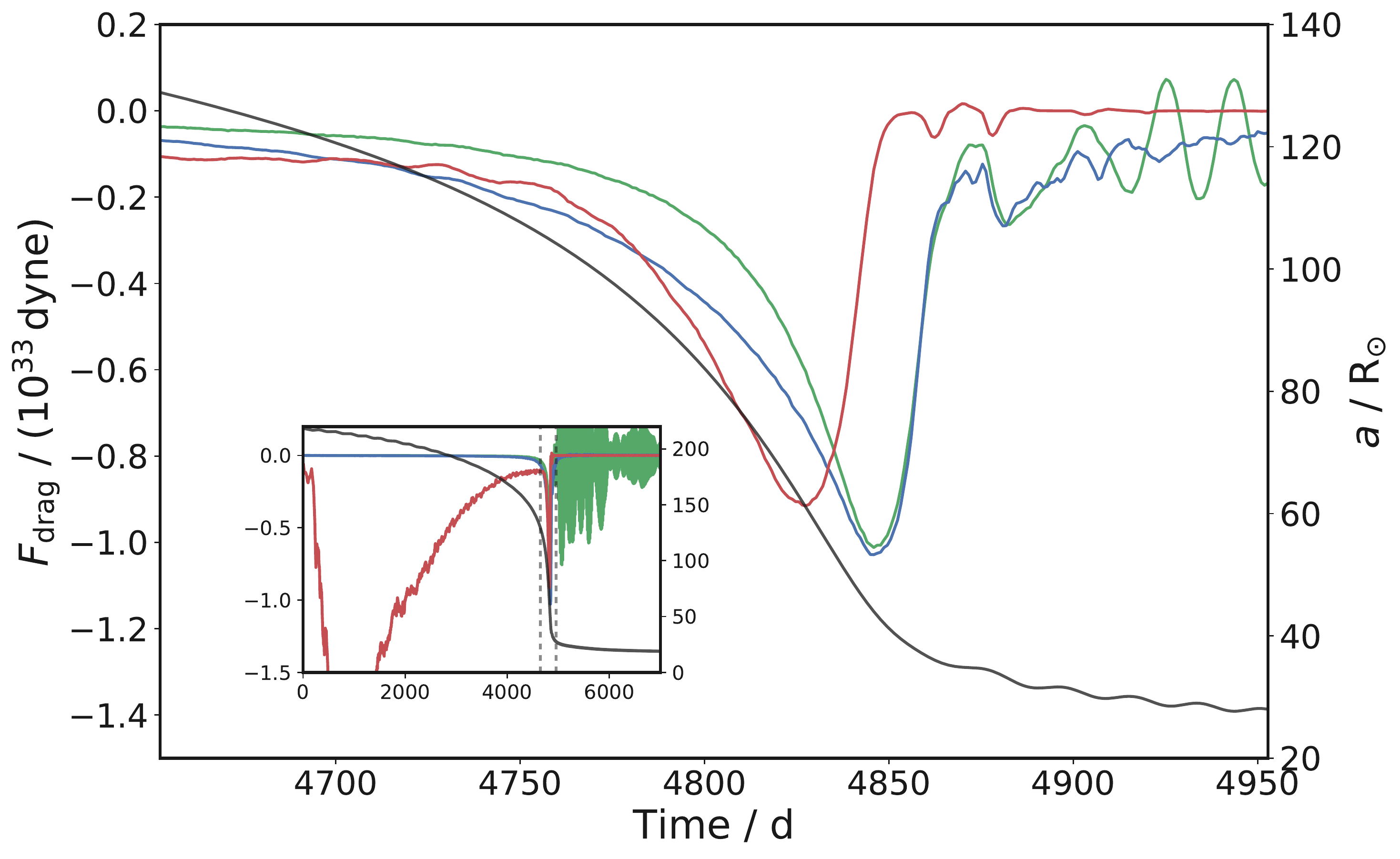}
    \caption{Gravitational drag force on the companion in the 1.1 million particle simulation, calculated with different approximations: summing interactions with nearby gas particles (blue line), by the observed loss of angular momentum from the orbit (green line; Eq.~\ref{eq:jdotdrag}) and by the analytical form of the drag force (red line; Eq.~\ref{eq:ana_drag}). The orbital separation is plotted in black. The inset is a zoom out over the entire simulation, for reference. Dashed black lines in the inset show the time range of the main plot, which is the time range during which the system is undergoing fast inspiral.}
    \label{fig:drag}
\end{figure}

The gravitational drag has been represented by several authors, with some variability of form \citep[e.g. see][]{shima1985hydrodynamic}, as
\begin{equation}
F_\text{drag} \approx \xi \pi R_\text{a}^2 \rho v_\text{rel}^2,
\label{eq:ana_drag}
\end{equation}
where $\xi$ is a factor dependent on the Mach number, $R_\text{a}=2GM_2/(v_\text{rel}^2+c_s^2)$ is the accretion radius\footnote{We use the form adopted for subsonic motion, typically observed in common envelope inspirals.}, $M_2$ is the mass of the companion, $c_s$ is the local sound speed, $\rho$ is the density around the core and $v_\text{rel}$ is the relative velocity of the core and the surrounding gas. The parameter $\xi$ is greater than 2 for supersonic motion and less than 1 for subsonic motion \citep{shima1985hydrodynamic} but, owing to difficulties in calculating this quantity on the fly, it is typically just held at unity.

We use Eq.~\ref{eq:ana_drag} to verify our simulation in the knowledge that the expression itself may have only limited validity in our situation. To use this expression we need to derive from the simulations the relative velocity of the companion and the density of surrounding material. First, we find the mean velocity vector of all gas within a given radius from the companion. In this case, the radius we use is the distance from the point mass particle to the centre of mass of the orbit. This varies from about 100\,\rsun\ to about 10\,\rsun\ as the orbit shrinks. The mean velocity vector of the gas is then projected onto the velocity vector of the companion, and the difference is found. The average density is similarly found within the same radius. An alternative method is to use a constant radius. This cannot be too small or else only the material caught in the potential well is encapsulated and cannot be too large or else in the late inspiral the core of the primary is included. \citet{iaconi2017effect} use a static radius of 20\,\rsun. Both methods have their benefits but, within the timespan of the dynamic inspiral, we find our method is more stable.

Any method which averages quantities over a volume around the companion includes substantial stationary gas caught in its potential well. This decreases the average velocity contrast, while it increases the density. Further, the accretion radius used in Eq.~\ref{eq:ana_drag} is also calculated with the the mean relative velocity and sound speed and is quite sensitive to slight changes in these quantities. Finally, Eq.~\ref{eq:ana_drag} is meant for use in a homogeneous medium, which is not the case in our simulations. In a common envelope, there is a relatively strong density gradient perpendicular to the motion of the core, along with an overdensity caught in its potential well, so the output of Eq.~\ref{eq:ana_drag} should be taken to be indicative at best.

There are two ways to calculate the drag force directly from the calculations. The first is via the change in angular momentum $\dot{J}$ of the point mass particles, with respect to the centre of the orbit by  

\begin{equation}
F_\text{drag} = \frac{\dot{J}}{r},
\label{eq:jdotdrag}
\end{equation}
where $r$ is the distance of the companion from the centre of mass. It should be noted that this only takes into account the component of the drag which is perpendicular to the radius from the centre of the orbit. Hence, while it is a reasonable approximation for circular orbits, if the cores are on an elliptical orbit this equation would yield a lower limit.

\begin{figure*}
	\includegraphics[width=\linewidth]{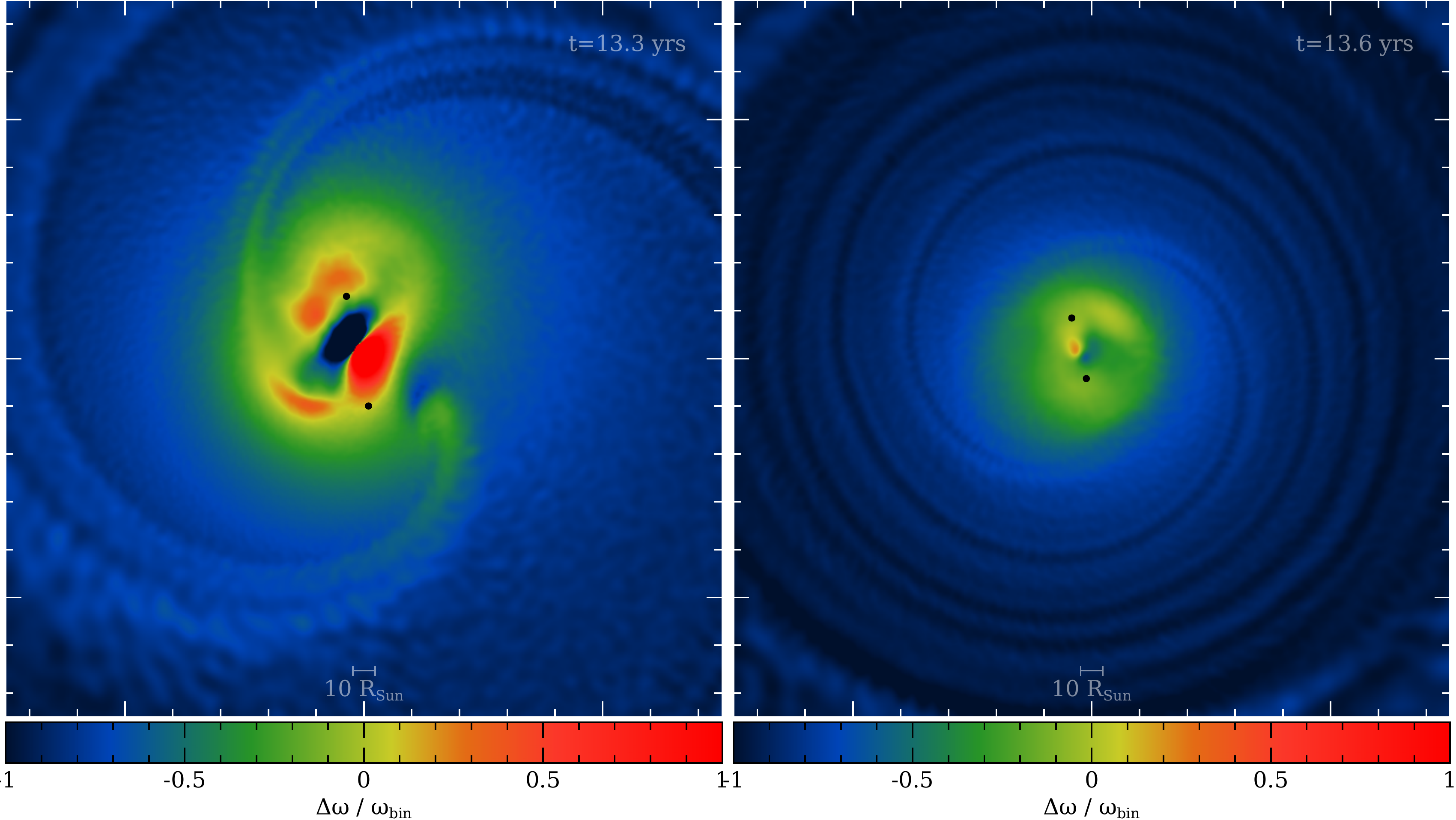}\\
    \caption{Slices in the XY-plane of the 1.1 million particle simulation, rendering the quantity $\Delta\omega/\omega_\text{bin}$, where $\Delta\omega \equiv \omega_\text{bin} - \omega_\text{gas}$, and $\omega_\text{bin}$ and $\omega_\text{gas}$ are the angular velocity of the point masses and the gas, respectively. The two snapshots are taken when the system is experiencing the most rapid reduction in orbital separation (left), and at the end of the dynamic inspiral (right). The frames have a side length of 1.5\,AU. Black dots represent the point masses.}
    \label{fig:corotation}
\end{figure*}

The second way to determine the dynamical friction in a simulation is to directly calculate the forces experienced by the cores as a result of other particles in the simulation. It is not useful to sum up the contributions from all particles, because the gravitational forces are dominated by the high density region surrounding the core of the giant. We select out a certain volume around the companion and determine the local gravitational attraction in this way. If the gas around the companion were spherically symmetric then there would be no net force on the particle. However, owing to the motion of the particle through the gas we see that the symmetry is broken by a high density wake that forms behind the companion. To calculate this force, we add up the gravitational forces between the companion and gas particles residing within the radius from the companion to the centre of mass of the orbit. This total force is then projected onto the velocity vector of the companion to give the dynamical friction of the gas on the companion particle. 

The analytical and two numerical forces are plotted in Fig.~\ref{fig:drag}, where the force is negative as it acts in the opposite direction to the orbital motion. In Fig.~\ref{fig:drag} we can see that the two numerical methods for calculating the drag force are comparable to within a factor of two. We recall that the torque calculation is a lower limit. However, the analytical force displays some differences though the peak magnitude remains within a factor of two of the blue line. Some variation can be obtained by changing the volume within which we measure the averaged quantities. The period of apparent strong analytical gravitational drag before the inspiral, seen in the inset of Fig.~\ref{fig:drag}, is an artefact of averaging over a large volume. The opposite effect occurs when the volume shrinks too much, because the force becomes entirely dominated by contributions of particles stuck in the potential wells of the cores, sending the drag force to zero earlier than it should. Also worth mentioning is the parameter $\xi$, which is dependent on the Mach number, and may vary the expected drag force by a factor of order unity.

\subsection{The Orbital Stabilisation Phase}
\label{ssec:theorbitalstabilisation}

The orbit of the two cores stabilises at around 20\,\rsun, independent of resolution. By the end of the high resolution simulation, the rate of decrease of the orbit has slowed down to about 0.5\,\rsun\,yr$^{-1}$, though this descent is still decelerating. 

In Fig.~\ref{fig:drag} we see that at some point there is a significant reduction of the drag force. The cause of this lower force is not so much the lower density of the surrounding gas, but that the gas has been dragged into corotation. Fig.~\ref{fig:corotation} shows two snapshots from the high resolution simulation, showing two cross-section slices of the quantity $(\omega_\text{gas} - \omega_\text{bin})/{\omega_\text{bin}}$, where $\omega_\text{gas}$ and $\omega_\text{bin}$ are the angular velocity of the gas and of the point masses around the centre of mass, respectively. The left-hand panel shows the time when $|\frac{\dot{a}}{a}|$ is a maximum, corresponding to the maximum rate of inspiral, while the right-hand panel shows the same quantities about a few months later, at the end of the dynamic inspiral. The gas surrounding the cores is rotating more uniformly in the second panel and is in approximate corotation, preventing the cores from losing further angular momentum to the envelope gas. This feature can also be seen in Fig.~\ref{fig:1e6angmom}, in which the angular momentum carried by the point mass particles remains approximately constant after the end of the dynamic inspiral.

What follows is complex and depends on how much of the envelope is unbound. Stellar structure dictates that for a giant star to depart from the giant branch and shrink, all but a very small amount of envelope must be lost. If the envelope is merely lifted by the common envelope inspiral, but not unbound, the orbit may stabilise while it has little gas within it. We could expect that if the envelope is ejected and the star departs the giant branch shrinking to within its Roche lobe, a small part of the envelope that was lifted but not ejected would return to form a low mass circumbinary disk. However, if the majority of the lifted envelope is not ejected, then a substantial amount of mass would return to the binary, resulting in a further interaction.

The time-scale of this return could be very small, though the ballistic calculation of \citet{kuruwita2016considerations} is only a lower limit because it does not account for gas pressure. Ultimately this question cannot be answered while we do not have a definitive answer to the energetics of the envelope and when it does or does not become unbound.

\subsection{Common Envelope Fallback} 
\label{ssec:fallback}

If most of the envelope is still bound after orbital stabilisation, we expect it to fall back rapidly. Simulations that use recombination energy tend to unbind much more of the envelope than those, such as ours, that do not \citep{nandez2015recombination}. It is likely that, had these simulations included recombination energy, they would be placed amongst the simulations that succeed in fully unbinding the entire envelope \citep{iaconi2018effect}. However, even leaving aside the controversy of how much recombination energy can be used to unbind the envelope \citep{ivanova2018use, soker2018radiating}, there are situations where the envelope would not be fully unbound even with recombination energy, such as those with initial masses $\gtrapprox 2$\,\msun\ \citep{nandez2016common}. In these cases substantial mass would fall back so that it is worth expending a few words on common envelope fallback gas.

\citet{kuruwita2016considerations} investigated how the bound part of the common envelope returns and whether it should form a disk thanks to the added angular momentum. They could not follow the entire outward and return journey of the gas, but using their setup could deduce that gas would return in a matter of months to a few years, and that much of this gas would return all the way to the core. However, their calculation did not account for the fact that gas, upon returning, collides with outflowing gas. Since SPH allows us to track envelope material far from the central binary, we can investigate the dynamics of this gas over a longer time-scale. This allows us to take another look at the idea of an envelope fallback event, beyond what was done by \citet{kuruwita2016considerations}.

\begin{figure}
	\includegraphics[width=\columnwidth]{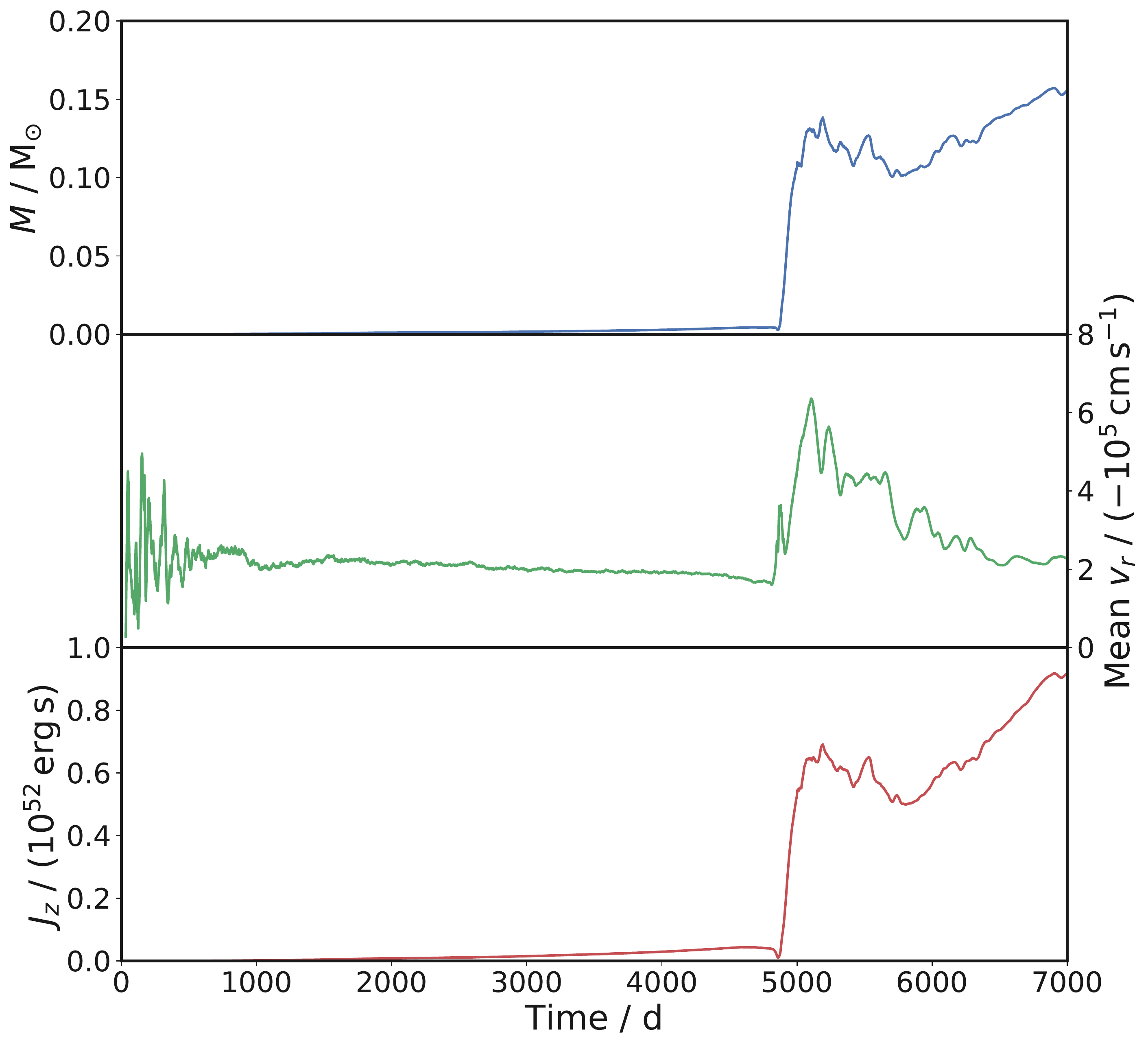}
    \caption{Fallback mass (blue), the average radial velocity of this mass (green), and the angular momentum tied up in this mass (red) as functions of time.}
    \label{fig:fallback}
\end{figure}

\begin{figure}
	\includegraphics[width=\columnwidth]{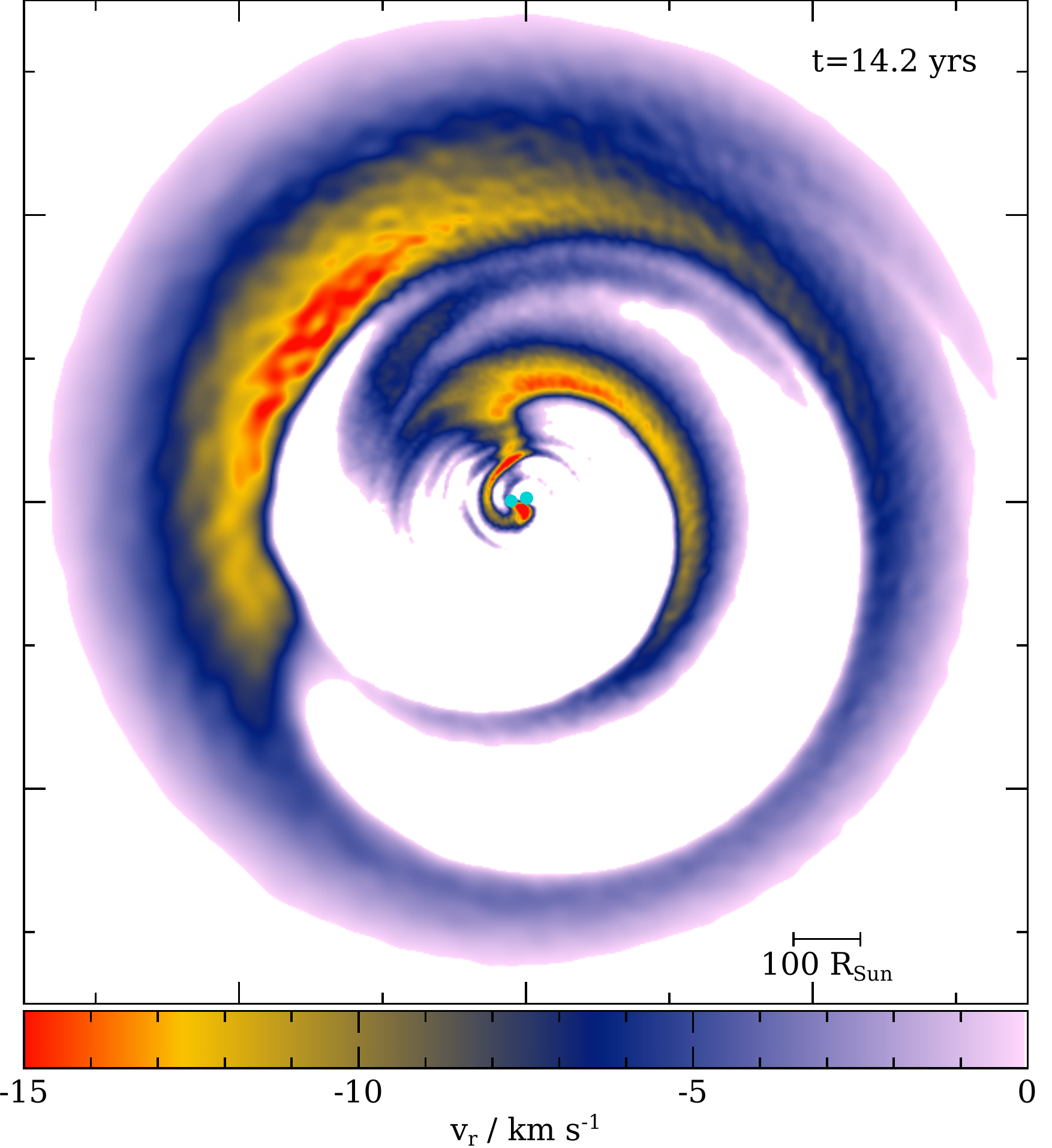}
    \caption{Distribution of gas displaying a negative radial velocity at t = 5188\,d for the 1.1 million particle simulation, when the most fallback mass was present in the simulation. Material is rendered by its \textit{negative} radial velocity value. Material with a positive radial velocity is not rendered. The frame is 7\,AU on each side.}
    \label{fig:fallbackslice}
\end{figure}

We consider SPH particles to be falling back if they have a negative radial velocity component with respect to the centre of mass and if they are located outside of a spherical volume encompassing the two stars. The total mass (blue line), angular momentum (red line), and the mean radial velocity (green line) of these particles are plotted in Fig.~\ref{fig:fallback}. Before the dynamic inspiral, there is very little material meeting our fallback criterion, as expected. During the inspiral, material is very rapidly evacuated from the region around the binary, catching up and colliding with material that was ejected previously. However, at the end of inspiral this evacuation stops and the material is able to fall back onto the central binary. The distribution of fallback mass is shown in Fig.~\ref{fig:fallbackslice}, and can be seen to be approximately circular at larger radii.

None of this fallback material returns to interact directly with the binary. Instead it impacts with more gas, flowing out from the region around the binary. It is possible that the region immediately surrounding the binary would become almost completely evacuated at some point, allowing future fallback material to return to the very inner portions of the system. However, we have not been able to follow the simulation for long enough to see this. 

\section{Planetary Nebulae and the Shape of the Common Envelope}
\label{sec:planetarynebulae}

At least one in five planetary nebulae (PNe) are ejected common envelopes \citep{miszalski2009binary}, and the symmetry axis of post-CE PNe with well characterised binaries can be shown to align with the orbital axis \citep{hillwig2016observational}. The shapes of the approximately 40 known post-CE PNe are interestingly inhomogeneous, although some common features, such as bipolarity, are observed in many cases \citep{de2013binary}. However, there is ultimately no obvious set of features that identifies the origin of a given PN as from a common envelope interaction.

PNe derive their shapes from the interaction of the fast and tenuous post-asymptotic giant branch (AGB) wind with a previous slow and massive AGB wind \citep{balick2002shapes}. The AGB slow wind creates a mould into which the fast wind expands. The common envelope provides gas distributions with an equatorial density enhancement. These are relatively small and dense compared to those generated by a slower mass-loss process. It is useful to determine the key differences between PN generated from post-CE distributions and those generated by single stars or other binary phenomena.

\citet{garcia2018common} have recently demonstrated how a common envelope density distribution, generated by the simulation of \citet{ricker2012amr}, can lead to a bipolar PN. They mapped the gas distribution at the end of common envelope simulation of  \citet{ricker2012amr} into a 2D plane, where only a quarter of the domain was simulated assuming two axes of symmetry, one along the orbital plane and one perpendicular to it. The input gas distribution from the common envelope simulation was taken after 56.7\,d of common envelope evolution, when the binary separation between the 0.36\,\msun\ degenerate core and 0.6\,\msun\ companion was approximately 9\,\rsun. A further 0.69~\msun\ of gas mass is in the ejected envelope. \citet{garcia2018common} modelled the central binary as a point with an effective temperature of 29\,000\,K, corresponding to a star with a wind velocity of 600\,km\,s$^{-1}$ (though they also performed two simulations with faster and slower winds). By evolving their simulation for $10^4$\,yr, they demonstrated that forming a bipolar nebula is, as predicted, a natural consequence of blowing a spherical wind into a toroidal gas distribution. Their simulations offer compelling evidence that connecting the common envelope with bipolar nebulae is a path certainly worth further study. 

\citet{frank2018planetary} used instead the simulation described here as a starting point for a PN simulation. Below we give further information as to the shape of the common envelope into which the post-AGB fast wind expands and compare their work further to that of \citet{garcia2018common}.

\subsection{The common envelope's gas distribution at the onset of the post-AGB fast wind}
\label{ssec:gasdistribution}

\begin{figure*}
	\includegraphics[width=0.75\linewidth]{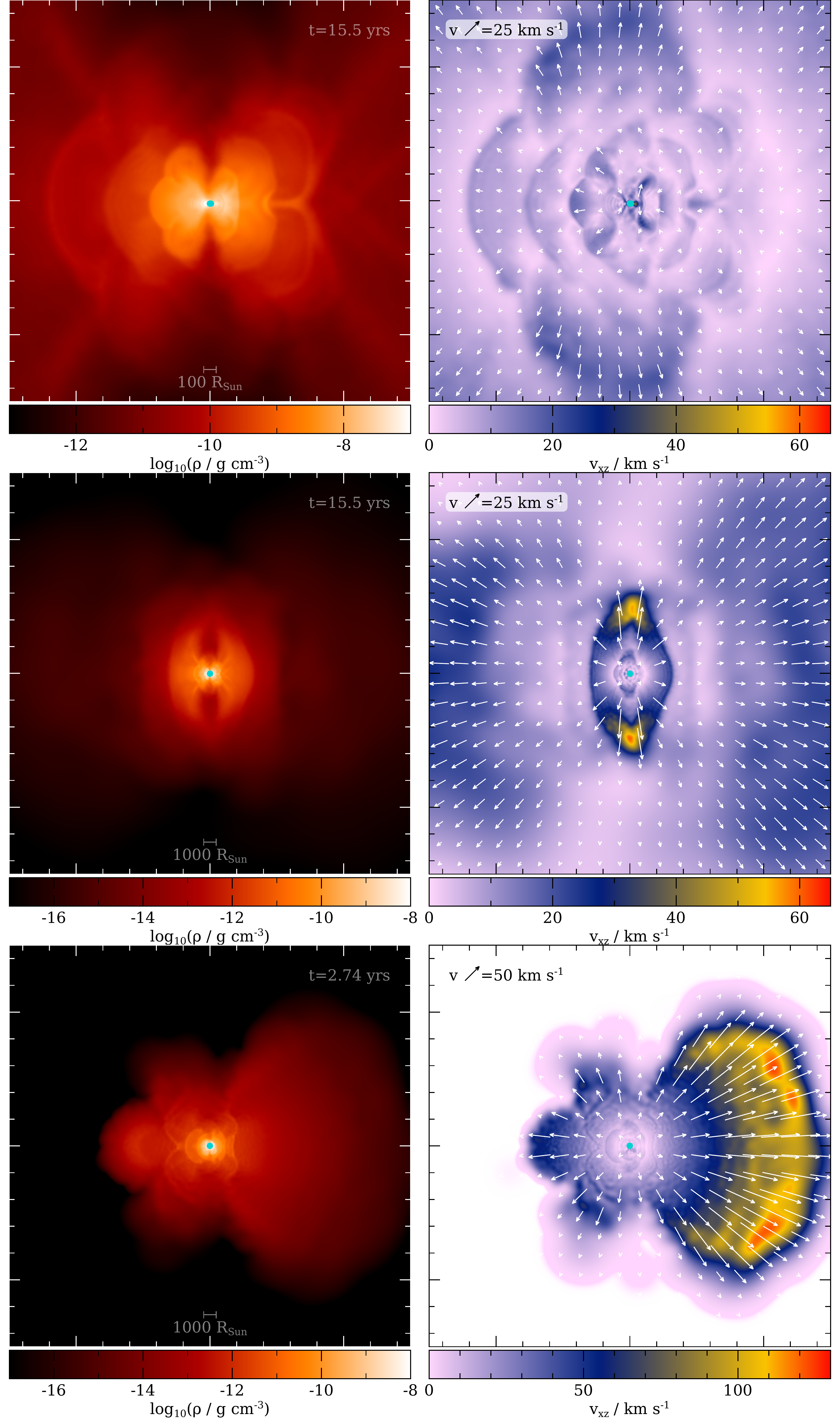}\\
    \caption{All frames are slices in the XZ-plane, rendered in density (left column) and velocity (right column). Top and middle rows display the our 1.1 million particle simulation from this paper, and differ by the scale: 15\,AU per side for the top row and 150\,AU for the middle row. The bottom row displays a simulation started with a smaller orbital separation of 100\,\rsun, but that is otherwise identical, and has side length of 150\,AU to match the middle row. Note the different scales of the colour bars.}
    \label{fig:frames}
\end{figure*}

The binary star considered by our simulation differs from that of \citet{ricker2012amr}. Our primary star is a 0.88\,\msun\ giant with a 93\,\rsun\ radius, while they have a 1.05\,\msun\ giant with a 31\,\rsun\ radius. Therefore their giant, is more bound and is more similar to a lower RGB star than the more extended AGB star that forms a PN. The more bound nature of their giant leads to a faster evolution over all and a more compact, denser common envelope ejection. Additionally, we evolved our simulation for longer, particularly before the inspiral. Despite the fact that both simulations start the companion near the beginning of Roche lobe overflow, our simulation spends a much longer time before inspiral, and more gas is launched out of the $L_2$ and $L_3$ points before the common envelope ejection takes place. This provides us with a more extended common envelope gas distribution into which the post-AGB wind expands.

In addition, the shape of the extended common envelope is greatly influenced by the interaction of the Roche lobe overflow with the outflow from the common envelope inspiral. During the Roche lobe overflow phase, gas leaves the system primarily through the second and third Lagrange points (see Fig.~\ref{fig:slices}). There is almost no unbinding of material in this phase (Fig.~\ref{fig:218sepbound}). The following fast inspiral phase interacts with the Roche lobe overflow ejecta. This distribution, shown both in the final row of Fig.~\ref{fig:slices} and in the top and middle frames of Fig.~\ref{fig:frames}, is toroidal. The polar regions in the distribution are approximately one to two orders of magnitude less dense than the material immediately surrounding it on the equatorial plane. The velocities of the ejecta are also considerably greater in a direction perpendicular to the orbital plane. Hence the relative difference in density continues to increase as the distribution expands.

The bottom row of Fig.~\ref{fig:frames} shows a slice of a simulation identical to those we have presented here, except that the initial orbital separation was set to 100\,\rsun. For more details on this simulation, see \citet{iaconi2017effect}. All of the slices in Fig.~\ref{fig:frames} are produced at 1000\,d after the beginning of the fast inspiral in their respective simulations. By comparing the middle and bottom row panels, we see that the gas distributions depend heavily on whether the pre-inspiral phase was simulated or not. Specifically, the Roche lobe overflow phase appears to promote symmetry in the final distribution. The bottom row frames show that there is a large lobe to the right of the frame which is rapidly expanding. There is a hint of the funnel regions of low density above and below the orbital plane but it is certainly not as distinct as the middle panels. {\it We suggest that the amount of mass loss before the inspiral makes a difference to the shape of the ensuing planetary nebula.} We therefore predict that differences in the nebular shape are the results of more or less stable pre-common envelope mass transfer, likely induced by a larger or smaller companion mass. It is worth noting that some of the morphological differences between the middle and bottom panels of Fig.~\ref{fig:frames} may be a result of having a non-rotating primary, which could plausibly lead to greater asymmetry in the bottom panels.

\subsection{The SPH-AMR PN Simulation}

In our high resolution simulation, all particles have an equal mass of $4.54 \times 10^{-7}$\,\msun. Typical post-AGB winds release mass on the order of $10^{-8}$\,\msun\,yr$^{-1}$. As \textsc{Phantom} currently does not support unequal mass particles, we are unable to model both the common envelope and the spherical wind in the same simulation. The 3D post-CE gas distribution has therefore been mapped into a series of nested grids using \textsc{Splash} \citep{price2007splash} and used as input to the 3D AMR grid code AstroBEAR \citep{cunningham2011astrobear}. Preliminary results of this PN simulation have been presented by \citet{frank2018planetary}. 

In summary, the AstroBEAR domain is 16\,000\,\rsun$^2$\ $\times$ 128\,000\,\rsun\ in size, mapped with seven levels of refinement. Two spherical wind cases are simulated. In one, a spherical wind with a mass-loss rate of $6.4 \times 10^{-4}$\,\msun\,yr$^{-1}$ is injected after a quiescent period of 500\,d, while in the other a $3.2\times 10^{-5}$\,\msun\,yr$^{-1}$ mass loss rate spherical wind is injected after a quiescent period of 6000\,d. Both winds have speeds of 300\,km\,s$^{-1}$. The simulations are each only run for approximately 1000\,d after the injection of the wind, which is much less than for the simulations of \citet{garcia2018common}. The wind is injected into the simulation at a radius of 46.9\,\rsun. A 1\,\msun particle is placed at the core and replaces the binary. This mass is also the only source of gravity in the simulation; no self-gravity is applied. Strong hydrodynamic collimation occurs in both these simulations, as well as in the simulation of \citet{garcia2018common}, due to the narrow, evacuated funnel that can be observed in Fig.~\ref{fig:frames}, upper left frame. We also note that the simulations of \citet{frank2018planetary} with the lower wind momentum display a strong degree of asymmetry between the upper and lower collimated lobes, with one much smaller than the other. 

Needless to say, these are early days for hybrid common envelope PN simulations such as these, especially because neither of the two sets of simulations have used a CE gas distribution that is unbound. This means that the envelope velocity field is likely not properly reproduced.

\section{Summary and Conclusions}
\label{sec:conclusions}

We have presented a set of hydrodynamic simulations of common envelope evolution aimed at extending the purely dynamical inspiral phase to include the previous, Roche lobe overflow phase as well as part of the post-inspiral phase. We also considered the post-common envelope, nebular phase. These simulations were carried out with the newly released smoothed particle hydrodynamics code, \textsc{phantom}, and are based on an initial model of a 0.88\,\msun, RGB primary star with a radius of about 90\,\rsun, and a 0.6\,\msun\ companion \citep{passy2012simulating}, placed at an initial orbital separation of 218\,\rsun, which is tuned to trigger Roche lobe overflow. Our main results are the following:

\begin{enumerate}
\item Our {\it unstable Roche lobe overflow} lasts for a short time (between 6.4 and 12.7\,yr, depending on resolution and 18\,yr if the giant is corotating with the orbit). From trends with primary star rotation, stability and resolution, we conclude that, in nature, this phase should last longer. Although we are unable to ascertain the length of the time between the onset of Roche lobe overflow and the fast inspiral with any certainty, we exclude that this phase is longer than a few centuries for binary parameters similar to those we have modelled.
\item By examining the ejection of mass and angular momentum, we are able to see that the orbital evolution of the system responds in a way that is expected analytically. The mass transfer rate during the Roche lobe overflow phase is aligned with analytical predictions to within a factor of a few. This gives confidence that improvements in the simulations may successfully and reliably model the Roche lobe overflow phase preceding a common envelope and soon lead to a better understanding of the stability criteria.

\item The {\it flow ejected via the $L_2$ and $L_3$ Lagrangian points} remains mostly bound: only 10$^{-3}$\,\msun\ are unbound with a further $7.8 \times 10^{-2}$\,\msun\ remaining bound. If unobstructed, fallback of this material would form a disk at a radius of about 130\,\rsun\ on a time-scale of at least 20\,yr. In our case, the unbound outflow associated with the subsequent dynamic inspiral disrupts this structure, in a manner similarly observed by \citet{macleod2018bound}. However, it is possible that, if the $L_2$/$L_3$ outflow phase were longer (mass transfer more stable), it may be followed by a weaker common envelope ejection, so that the disk  structure formed during Roche lobe overflow may remain to become a long-lived circumbinary disk, as has been observed around post-AGB binaries with periods between 100 and 1500\,d \citep{vanwinckel2009post}. 

\item \citet{pejcha2016binary} simulated an $L_2$ flow in a binary that was not allowed to change its orbital separation. While they concluded that the $L_2$ outflow becomes unbound for mass ratios (accretor and donor) in the range 0.06--0.8, this is at odds with both the results of our simulations and those of \citet{macleod2018bound}.

\item The {\it gravitational drag felt by the cores during the fast inspiral} is within a factor of two of that calculated analytically using quantities from the simulation, as was also concluded for planetary mass companions by \citet{staff2016hydrodynamic}. This consistency exists despite the many approximations and assumptions and it increases our confidence in the inspiral timescale being a reasonable approximation of reality. The {\it end of the inspiral} is brought about by gas being brought into co-rotation with the binary orbital motion. 

\item The low, medium and high resolution {\it simulations unbind 33, 16 and 11 per cent of the gas mass, respectively} (calculated without including internal energy). Hence there is a trend to unbind less mass as resolution increases \citep{iaconi2017effect}. Without having stabilised the star in the corotating frame, we are unable to definitively state what effect a rotating star has on the amount of unbound mass. 

\item During the early stages of the {\it self-regulated inspiral} following the dynamic inspiral, as much as 30 per cent of the gas displays negative radial velocities. This fallback material is typically within 500\,\rsun\ of the central binary though, if the simulation were to be run for longer, material at larger radii would start to fall back. 

\item The final gas distribution of our simulation is toroidal, displaying a strong density contrast between the orbital plane and the polar directions. The common envelope ejection following the disk formation develops high velocities in the direction perpendicular to the orbital plane. However, this perpendicular  ejection contains relatively little mass. It is not this mass that becomes visible as a bipolar planetary nebula later on. The highly evacuated polar funnels provides strong density contrast and powerful hydrodynamic collimation, which can be seen in the nebular simulations of \citet{frank2018planetary}, carried out with the output of this simulation.

\item By comparing our simulation to an identical one, except for the fact that the companion was placed at the giant's surface, and in which the dynamic inspiral develops immediately, suggests that simulating the Roche lobe overflow phase results in a far more symmetric distribution of gas, due to the regulating effect of the disk on the subsequent morphology. This is something that would impact the shape of subsequent PNe. This in turn suggests that the length of the Roche lobe overflow phase or, in other words, {\it the degree of stability of the mass transfer, may explain the range of PN shapes of common envelope PN.}
\end{enumerate}




\section*{Acknowledgements}

TR acknowledges financial support through the Macquarie University Research Excellence scholarship associated with Future Fellowship grant to O. De Marco (FT120100452). OD and DP acknowledge financial support through the Australian Research Council Future Fellowship scheme (FT120100452 and FT130100034, respectively). CAT thanks Churchill College for his fellowship and Macquarie University for supporting a visit. This work was supported by access to the swinSTAR supercomputing facility at Swinburne University of Technology.




\bibliographystyle{mnras}
\bibliography{mnras_bib} 



\bsp	
\label{lastpage}
\end{document}